\documentclass{egpubl}
\usepackage{eg2022}

\usepackage[T1]{fontenc}
\usepackage{dfadobe}  
\usepackage{xcolor}

\biberVersion
\BibtexOrBiblatex
\usepackage[backend=biber,bibstyle=EG,citestyle=alphabetic,backref=true]{biblatex} 
\electronicVersion
\PrintedOrElectronic

\ifpdf \usepackage[pdftex]{graphicx} \pdfcompresslevel=9
\else \usepackage[dvips]{graphicx} \fi

\usepackage{egweblnk} 
\usepackage[english]{babel}
\usepackage{amsmath}
\usepackage{amssymb}

\newtheorem{theorem}{Theorem}[section]
\newtheorem{lemma}[theorem]{Lemma}

\title{Distortion Reduction for Off-Center Perspective Projection of Panoramas}

\author[Chi-Han Peng \& Jiayao Zhang]
{\parbox{\textwidth}{\centering Chi-Han Peng$^{1}$ and Jiayao Zhang$^{2}$}
        \\
{\parbox{\textwidth}{\centering $^1$National Chiao Tung University\\
         $^2$King Abdullah University of Science and Technology}
}
}

\begin{document}

\teaser{
 \includegraphics[width=\linewidth]{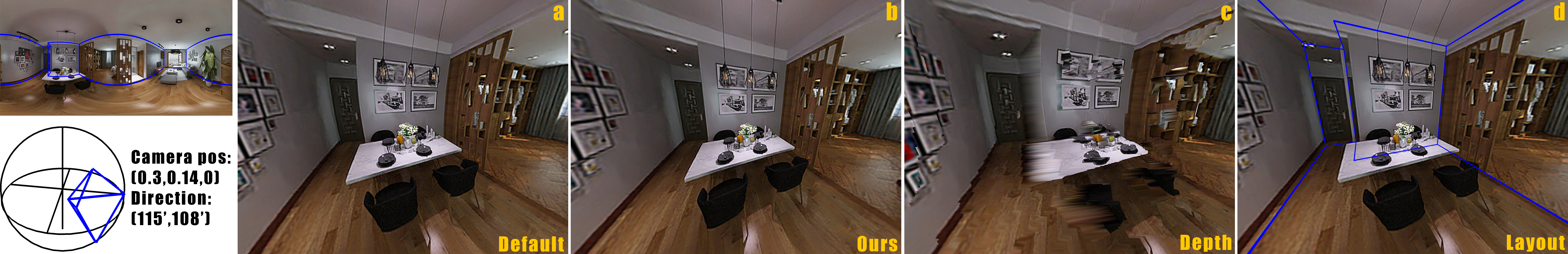}
 \label{fig:teaser}
 \centering
  \caption{(a) Off-center perspective projections of a panorama contain barrel distortions and break realism. (b) With our distortion-reduction measures, distortions are significantly reduced and the visualization is more like a genuine perspective projection in which linear lines in 3D remain linear in 2D. In comparison, augmenting the panorama with (c) depth or (d) room layouts do not lead to visually convincing results. Observe the broken lamp wires and doorway in (d). We show the panorama and room layout (predicted by LED2-Net~\cite{wang2021led2}) and the camera position and looking direction on the left. The panorama and the ground-truth depth are from the Structure3D dataset~\cite{Structured3D}. } 
\label{fig:teaser}
}

\maketitle
\begin{abstract}
A single Panorama can be drawn perspectively without distortions in arbitrary viewing directions and field-of-views when the camera position is at the origin. This is a key advantage in VR and virtual tour applications because it enables the user to freely "look around" in a virtual world with just a single panorama, albeit at a fixed position. However, when the camera moves away from the center, barrel distortions appear and realism breaks. We propose modifications to the equirectangular-to-perspective (E2P) projection that significantly reduce distortions when the camera position is away from the origin. This enables users to not only "look around" but also "walk around" virtually in a single panorama with more convincing renderings. We compare with other techniques that aim to augment panoramas with 3D information, including: 1) panoramas with depth information and 2) panoramas augmented with room layouts, and show that our approach provides more visually convincing results. 


\end{abstract}  

\section{Introduction}
\label{sec:introduction}

A spherical panorama stores incoming ray intensities toward a fixed camera point in all possible directions. Commonly, the directions are sampled on a sphere (centered at the point) in a 2D equirectangular format. Using just a single panorama, one can render accurate perspective projections from the camera point in arbitrary viewing directions and field-of-views through sampling strategies commonly known as the equirectangular-to-perspective (E2P) projection. In VR and virtual tour applications, this is known as the 3DoF (i.e., rotations along three axis) of a perspective camera in a virtual environment, albeit at a fixed position.

It is actually straightforward to render perspective projections from an {\em off-center} position, although the results would contain barrel distortions. One way is to map the panorama onto a spherical mesh through projective texture mapping and draw perspective views using standard perspective camera models such as OpenGL. An analytical form of the off-center E2P projection is described in Section~\ref{sec:E2P}. The rendering results (see the accompanying video for examples) are "intuitive" in general, i.e., scenery becomes bigger when the camera moves closer and the scenery moves to the right when the camera moves to the left, and vice versa. Parallax effect and occlusion on-and-off are lacking. However, by our observations, lacking them usually doesn't ruin the plausibility of the rendering, unless there are large variations in the depth disparity (e.g., having objects very close to the camera).

We opt for solving another major defect of the rendering - the {\em barrel distortions} (e.g., Figure~\ref{fig:teaser} (a)), which are inevitable when viewing a textured sphere from an interior position away from the center. The reason why barrel distortions break realism is that {\em straight lines in 3D no longer remain straight in the 2D view}, breaching a key assumption of perspective projections.

Such task is often called the novel view synthesis problem in visual computing. Powerful methods such as~\cite{xu2021layout} leverage deep learning (DL) models and comprehensive data to synthesize novel views of the same scene that extrude parallax effect and even contain scenery that was previously occluded. These DL-based methods either take panoramic videos as input (e.g., "one-shot 3D photography"~\cite{Kopf2020}), or need a dataset trained on such (\cite{xu2021layout}). In comparison, our method takes a lightweight approach to the problem (no DL training/inference nor video-sequence inputs are needed, and is very computationally light), with a focus on tackling barrel distortions.

We introduce two modifications to the standard E2P projection process. First, in order to preserve the linearity of vertically straight lines in 3D in the 2D view, we re-project the panorama to a {\em cylinder} so that all vertically straight lines in 3D, which are necessarily mapped to meridians on the sphere, would be projected to vertically straight lines in 3D again (i.e., the straight lines on the cylinder), which are guaranteed to be projected to straight lines in 2D. Second, we introduce computational {\em dolly-zoom} effects to find alternative camera position and field-of-view for the same viewing region that minimizes barrel distortions. In summary, the first measure effectively eliminates barrel distortions of vertical lines while the second measure reduces distortions in both horizontal and vertical directions to a degree.

The endgame for the novel view synthesis problem is to build an accurate 3D model of the scene. To our best knowledge, \cite{xu2021layout} is the only DL-based work that takes a single panorama as input for such a goal. However, their code is not yet available. Therefore, we opt for comparing to two common strategies to build a rough 3D model out of a single panorama: 1) panoramas with per-pixel depths (ground-truth or predicted by neural networks), and 2) augmenting panoramas with room layouts. In our experiments (see Section~\ref{sec:result}), we show that both strategies are inadequate to produce good quality rendering results.

The paper outline is as follows. In Section~\ref{sec:related_work}, we describe related work in the novel view synthesis problem based on panoramic inputs. In Section~\ref{sec:method}, we describe our two distortion reduction measures and provide a thorough analysis. In Section~\ref{sec:result}, we compare the renderings of off-center perspective projections of our method to vanilla E2P projections and common approaches to build 3D panoramas. We further compare to results of~\cite{xu2021layout} (by using panorama images taken from their paper) and discuss the pros and cons. Results on a large variety of panoramas taken from different datasets are shown. Finally, we conclude the paper in Section~\ref{sec:conclusion}.


\section{Related work}
\label{sec:related_work}

\subsection{Panoramic 3D modeling and datasets}

3D Modeling of indoor scenes based on spherical panoramic (also called "360°" in commercial settings) image inputs is a popular field in recent years. Key tasks include depth estimation~\cite{pintoreslicenet,albanis2021pano3d,wang2020bifuse,lasinger2019towards}, room layout estimation~\cite{lin2020deep,xu2021layout,yang2019dula}, object detection and segmentation~\cite{sun2021hohonet,xia2018gibson}, and more generally 3D reconstruction tasks such as registration of multiple panoramas~\cite{yang2020noise,chen2021sio}.

A number of panoramic image datasets have been produced to aid the research. Matterport3D~\cite{chang2017matterport3d} and Standford2D3D~\cite{armeni2017joint} are real-world large-scale RGB-D datasets, which provide panoramic views of diverse indoor scenes. They include various kinds of 2D and 3D semantics, meshes, and even video walkthroughs. Structure3D ~\cite{Structured3D} and SunCG ~\cite{zioulis2018omnidepth} dataset are synthetic datasets with richly decorated indoor scenes. They provide realistically rendered indoor RGB-D images and annotations of 3D structures. 3D60~\cite{zioulis2019spherical} is a collective dataset with 3 different modalities (color, depth and normal) and comprises of realistic and synthetic 3D datasets (Matterport3D, Stanford2D3D, and SunCG). Gibson~\cite{xia2018gibson} is a real-world dataset that includes high-quality RGB panoramas, global camera poses, and 3D meshes.

\subsection{Novel view synthesis}



Novel view synthesis is one of the core tasks in visual computing. We limit our scope to methods rely on panoramic image inputs. Layered Depth Images (LDI)~\cite{shade1998layered} and Multiplane Images (MPI)~\cite{zhou2018stereo} are used as image-based representations for novel view synthesis, but for large translations, they might lack sufficient information to render correctly. Multi Depth Panoramas (MDPs)~\cite{lin2020deep} and PerspectiveNet~\cite{novotny2019perspectivenet} comprise of multi-RGBD$\alpha$ images for high-quality and efficient novel view generations. Hedman et al.~\cite{hedman2018instant} input burst of aligned color-and-depth photos to generate 3D panorama, and their 3D effects could also interact with the scene. Gao et al.~\cite{gao2021dynamic} propose an algorithm to generate novel views from dynamic monocular videos. Xu et al.~\cite{xu2021layout} make the first attempt to generate a target-view panorama from one single source-view panorama with a large camera translation. Jin et al.~\cite{jin2020geometric} and Zeng et al.~\cite{zeng2020joint} leverage the geometric structure of a 360° indoor image for depth estimation.

Attal et al.~\cite{attal2020matryodshka} simultaneously learn depth and occlusions via a multi-sphere image representation, which could greatly handle occluded regions in dynamic scenes. With 46 input light field video, Broxton et al.~\cite{broxton2020immersive} present a system that is able to reproduce view-dependent reflections, semi-transparent surfaces, and near-field objects. Tobias et al. introduce OmniPhotos~\cite{bertel2020omniphotos} for quickly and casually capturing 360° VR panoramas, and improve the visual rendering quality by alleviating distortion using a novel deformable proxy geometry. Serrano et al.~\cite{serrano2019motion} present a device which enable head motion parallax in 360°video, thus tackled silhouettes and occlusions. Other works extend these approach to point clouds, Aliev et al.~\cite{aliev2020neural} present a point-based approach to generate novel views of the scene. Voxel grid-based methods such as DeepVoxels~\cite{sitzmann2019deepvoxels} encodes the view-dependent appearance of a 3D scene as 3D voxels. Implicit function-based methods such as Sitzmann et al~\cite{sitzmann2019scene} propose a continuous, 3D structure-aware scene representation that encodes both geometry and appearance. 

Overall, nearly all existing methods reply on input data that consists of multiple panoramas (taken at different camera positions), often as panoramic videos shoot either from a single moving 360° camera or an array of fixed (or even moving) cameras (capturing rigs). One  exception is~\cite{xu2021layout}, in which they rely on DL-based methods to synthesize novel views from just a single panorama. Our work is most similar to theirs.


\section{Method}
\label{sec:method}

We first describe the analytical form of the vanilla off-center E2P projection in Section~\ref{sec:E2P}. We then describe our two measures to reduce barrel distortions in Section~\ref{sec:cylinder} and~\ref{sec:dolly}. Each approach can be applied independently. We assume a system would apply both unless otherwise specified.

\subsection{Off-center E2P projection}
\label{sec:E2P}

\begin{figure}[t]
  \centering
  \includegraphics[width=1\linewidth]{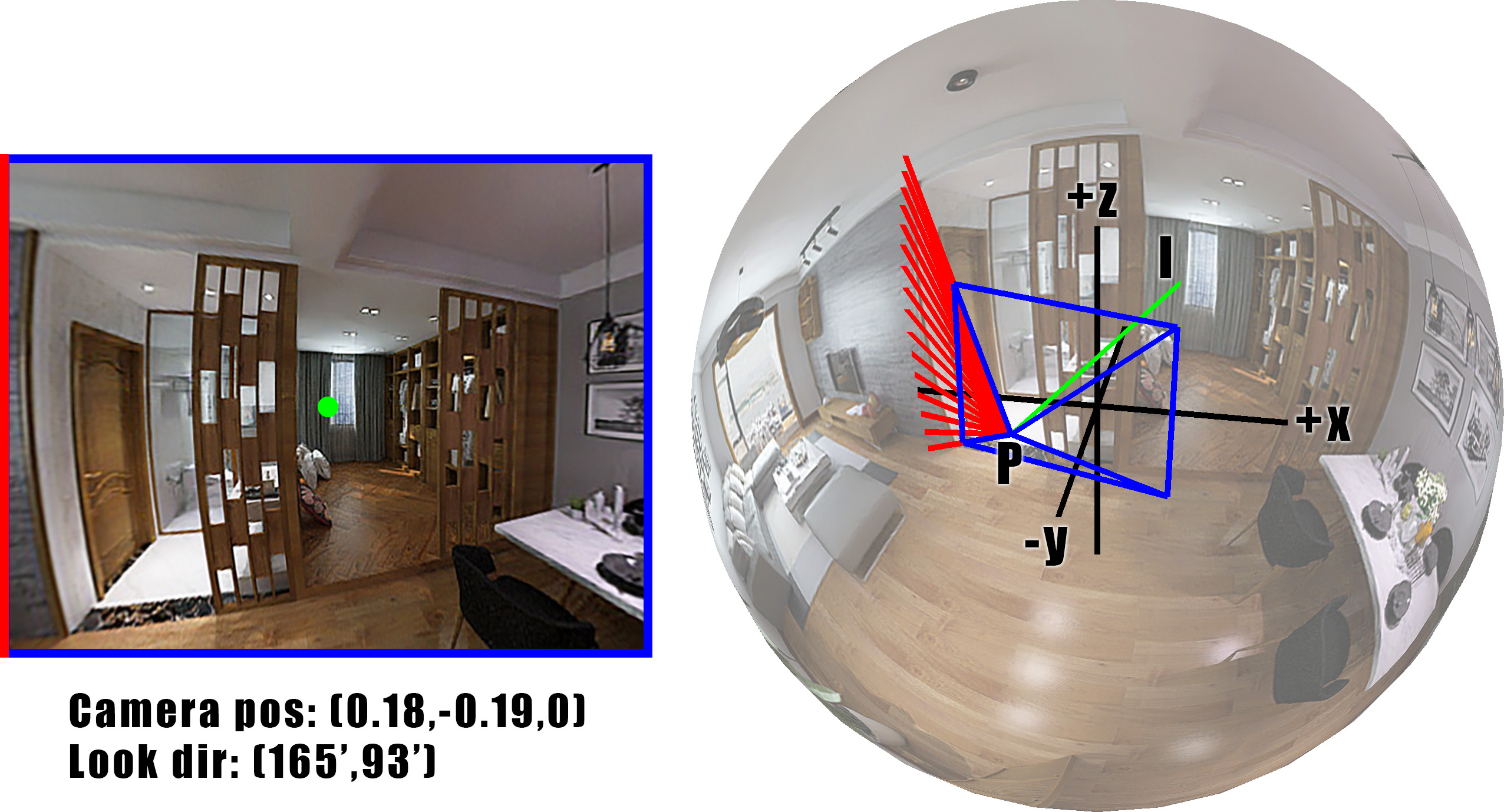}
  \caption{\label{fig:spherical_camera} An off-center E2P projection. $P$ is the camera center and $I$ is the intersection of the looking ray and the unit sphere. We use a right-handed system with +z direction at azimuth=0° and zenith=0° and +x direction at azimuth=0° and zenith=90°. The viewing pyramid is drawn in blue. Red lines show a column in the view plane that depicts a non-straight geometry in the 3D scene due to barrel distortions.}
\end{figure}

Without loss of generality, we assume the panorama is mapped to a unit sphere centered at the origin. We assume the perspective camera's position is $P = (p_x, p_y, p_z)$, looking direction is $Dir = (dir_x, dir_y, dir_z)$, and up direction is $Up = (up_x, up_y, up_z)$. $Up$ is orthogonal to $Dir$. We assume the image plane is rectangular and its local y-axis is aligned to the up direction. The dimensionality of the image plane can be described by two angles $fovx$ and $fovy$, i.e., the horizontal and vertical angles of the pyramid formed by the camera center and the view plane. Therefore, the width and height of the image plane are $width = near \cdot tan(fovx/2) \cdot 2$ and $height = near \cdot tan(fovy/2) \cdot 2$, respectively. $near$ is the distance between the camera position and the view plane. We denote the "left" direction as $Left = Up \times Dir$ by right-handed cross product. See Figure~\ref{fig:spherical_camera} for an example.

To retrieve the ray intensity (e.g., color or depth) at a normalized 2D position on the image plane, $(X,Y)$, $0 \le X,Y \le 1$, $X$ is left to right direction and $Y$ is top to bottom direction, we derive the corresponding 3D ray direction, $Ray = (ray_x, ray_y, ray_z)$, as:
\begin{equation*}
|Dir \cdot near + Left \cdot (0.5 - X) \cdot width + Up \cdot (0.5 - Y) \cdot height|,
\end{equation*}
which can be shortened as:
\begin{equation}
\label{eq:ray}
|Dir + Left \cdot (1 - 2 \dot X) \cdot tan(fovx/2) + Up \cdot (1 - 2 \dot Y) \cdot tan(fovy/2)|.
\end{equation}

To draw a perspective projection, for each 2D position $(X,Y)$ on the image plane (e.g., a pixel), our goal is to find the spherical coordinate, $(\theta,\phi)$, of the point $I$ on the unit sphere that intersects with the ray from the camera position to the point on the 2D image plane. Recall that $\theta$ is the zenith (angle from the +z axis), $\phi$ is the azimuth (angle of counterclockwise rotation along +z-axis from the +x axis), and the corresponding 2D coordinate in equirectangular projection can be trivially derived as $(\frac{\phi}{2\pi}, \frac{\theta}{\pi})$. 

We find the 3D coordinate of $I$ by solving $t$ in $I = P + t \cdot Ray$, $|I| = 1$. We have $a = ray_x^2 + ray_y^2 + ray_z^2$, $b = 2 \cdot (pos_x \cdot ray_x + pos_y \cdot ray_y + pos_z \cdot ray_z)$, and $c = pos_x^2 + pos_y^2 + pos_z^2 - 1$. $t$ equals $(-b + \sqrt{b^2 -4ac})/2a$ (i.e., we take the positive-signed solution). Finally, convert $I$'s 3D coordinate to spherical coordinate $(\theta,\phi)$.

Note that the above calculation can be done in the OpenGL rendering pipeline by setting up a unit spherical model centered at origin with the panorama as the 2D texture and equirectangular coordinates as UV coordinates, and drawing the 2D view defined by a view frustum defined by the aforementioned view pyramid cut through the near plane. A pixel shader is used to calculate the UV coordinates per pixel.

\subsection{Cylindrical projection}
\label{sec:cylinder}

\begin{figure*}[t]
  \centering
  \includegraphics[width=1\linewidth]{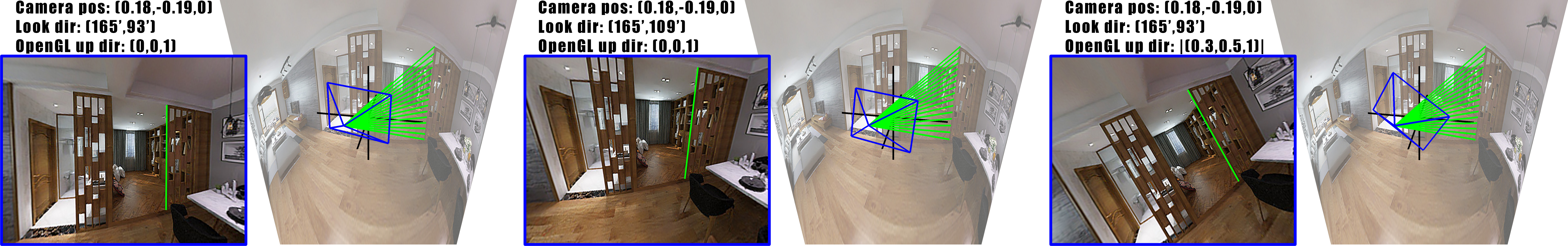}
  \caption{\label{fig:cylindrical_cameras} Off-center cylindrical projections. (a): A rendering using the same camera position and looking direction as in Figure~\ref{fig:spherical_camera}. All vertical features in the scene are now drawn as straight lines. (b): After the camera is pitched downward. The vertical features are still drawn as 2D straight lines, but not vertical. (c): After the camera is rolled clockwise. Again, vertical features are drawn as 2D straight lines.} 
\end{figure*}

Modern 360° cameras have strong image stabilization features to ensure that the shoot panoramas are nearly in upright position, i.e., the -z direction in 3D or the +y direction in equirectangular projection are aligned to the direction of gravity. There also exist algorithms to further transform a panorama to make it in upright position (e.g., the "camera rotation pose alignment" in~\cite{SunHSC19}). This means that vertical features in 3D in the scene, such as wall corners, are nearly always mapped to vertical lines in the panorama. However, this doesn't mean that they are mapped to vertical lines in 3D through the equirectangular projection. Instead, they are mapped to {\em meridians on the unit sphere}, which are curved in 3D and are perspectively drawn as curved lines in 2D except when the camera position is at the origin.

We propose a simple solution to ensure that vertical features in the scene are perspectively drawn as straight lines in 2D even when the camera is not at the origin: {\em projecting the panorama to a cylinder}. We choose the cylinder to be of radius $1$, centered at the origin, in the "upright" position (i.e., axis is aligned to the z axis), and of an infinite height. We now have off-center equirectangular-to-cylinder (E2C) projection as follows.

Equation~\ref{eq:ray} again describes the 3D ray direction, $Ray$, of a 2D position $(X,Y)$ on the image plane of the camera. Its intersection point with the cylinder, $I^c$, is calculated by solving $t'$ in $I^c = P + t' \cdot Ray$, $|I^c_x+I^c_y| = 1$. We now have $a' = ray_x^2 + ray_y^2$, $b' = 2 \cdot (pos_x \cdot ray_x + pos_y \cdot ray_y)$, and $c = pos_x^2 + pos_y^2 - 1$. $t'$ equals $(-b' + \sqrt{b'^2 -4a'c'})/2a'$. Next, convert $|I^c|$ to spherical coordinate $(\theta,\phi)$, and use it to sample a ray intensity in the panorama. 

We have the following lemma:

\begin{lemma}
\label{lemma:cylindrical}
A vertical feature in 3D in the scene is perspectively drawn as a 2D straight line under a cylindrical projection.
\end{lemma}

\begin{proof}
Recall that a vertical feature in 3D is necessarily drawn as a vertical line in the 2D panorama under equirectangular projections, which is then mapped to a part of a meridian on the sphere. Trivially, a cylindrical projection maps every meridian (and parts of it) on the sphere to a vertical straight line on the cylinder. Finally, recall that perspective projection preserves the linearity of any 3D straight lines in 2D perspective views. 
\end{proof}

Note that Lemma~\ref{lemma:cylindrical} applies for cameras with arbitrary rolls, pitches, and yaws, not just "upright" cameras with a left direction perpendicular to the z axis. See Figure~\ref{fig:cylindrical_cameras} for an example.

To realize the above calculation in the OpenGL rendering pipeline, replace the spherical model with a cylindrical model. The model is textured using the panorama as the 2D texture and "normalized" equirectangular coordinates, which are based on the spherical coordinates of the normalized 3D point positions, as UV coordinates.

In summary, Lemma~\ref{lemma:cylindrical} states that any vertical features in the 3D scene are guaranteed to be drawn as straight lines in the 2D perspective views. However, the inverse is not necessarily true - {\em everything} depicted on a column in the panorama would remain straight in 2D perspective views, no matter they are genuinely vertically aligned in 3D or not. In practice, we observe that the renderings largely remain smooth and intuitive.

\subsection{Computational dolly-zoom effect}
\label{sec:dolly}

In cinematography, a "dolly-zoom" effect refers to the act that the field-of-view (FOV) angle is continuously narrowed while the camera is moving away from an object in the scene, or vice versa, in a calibrated manner such that the object would appear at the same size on the 2D frame during the movement. The goal is to create so called "perspective distortions" in which the relative sizes of other objects in the scene change w.r.t. to the size of the particular object.

Recall that in perspective drawing of panoramas, in general, the amount of distortions is proportional to the distance between the camera position and the origin (zero when they coincide). Therefore, our main idea is to leverage dolly-zoom principles to draw roughly the same subset of the panorama in the image plane but from a camera position that is {\em as close to the origin as possible}. We call the new camera position, $P^h$, the heuristic solution to the computational dolly-zoom problem. 

\begin{figure}[t]
  \centering
  \includegraphics[width=1\linewidth]{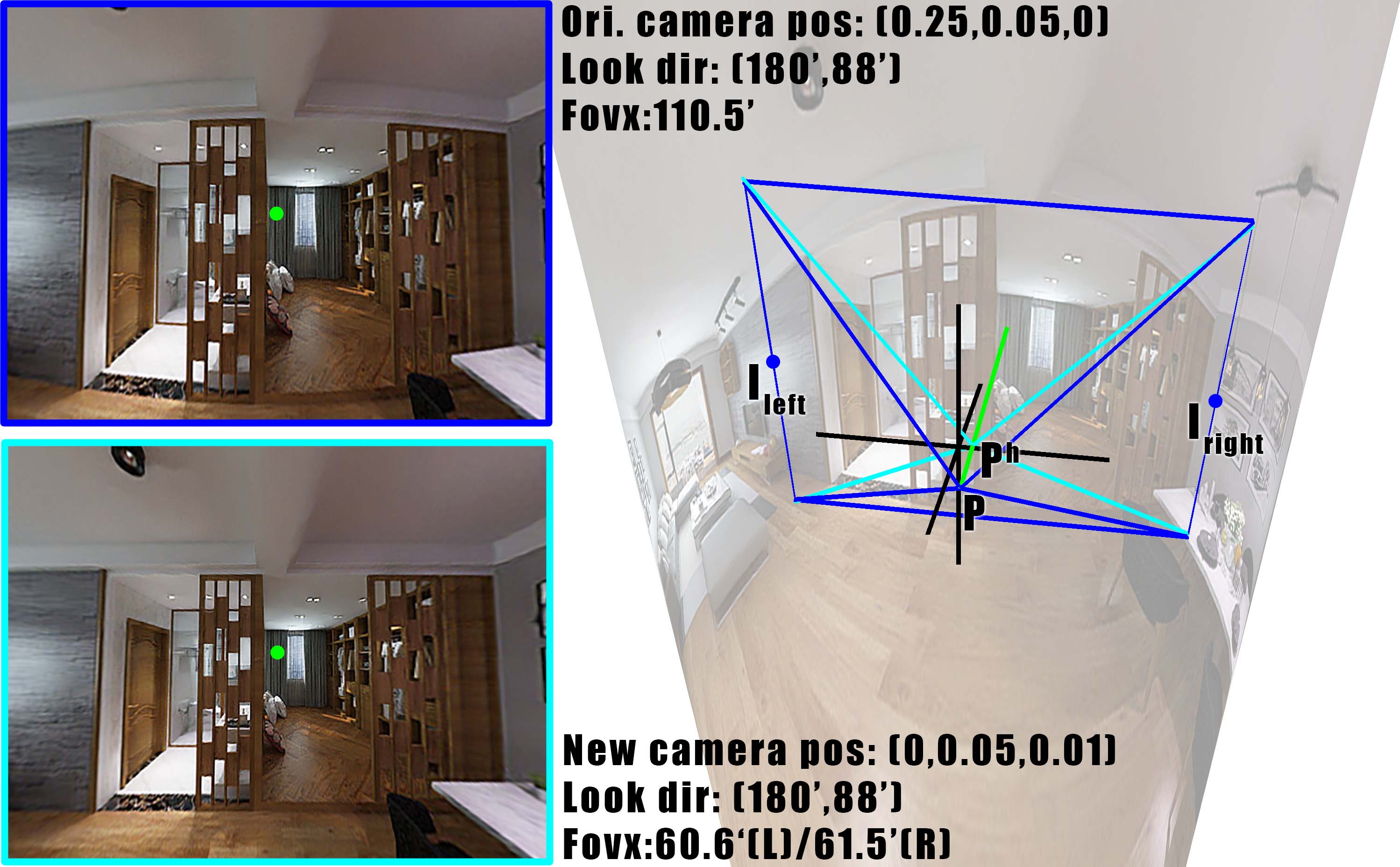}
  \caption{\label{fig:dolly1} Inspired by the dolly-zoom effects in cinematography, we find an alternative camera position and FOV angles setting that with which the perspective camera draws roughly the same subset of the panorama but with less barrel distortions. The original camera is shown in blue and the adjusted camera is shown in cyan. Observe that the two drawn images have roughly the same boundary and center. Just that the images are stretched in different ways.} 
\end{figure}

To elaborate, $P^h$ is the solution to the following optimization problem:
\begin{equation}
\begin{aligned}
& \underset{t}{\text{argmin}}
& & |P^h| \\
& \text{subject to}
& & P^h = P + t \cdot Dir
\end{aligned}
\end{equation}
where $P$ is the original camera position and $Dir$ is the looking direction. Given the new camera position $P^h$, we solve the new FOV angles as follows. First, we denote the "left-middle" and "right-middle" viewing rays, $Ray_{left}$ and $Ray_{right}$, as the rays from $P$ toward the left-middle and right-middle points of the original image plane. See Figure~\ref{fig:dolly1} for an example. Next, we find the intersections of $Ray_{left}$ and $Ray_{right}$ to the sphere or cylinder (depending on which projection scheme is used), denoted as $I_{left}$ and $I_{right}$, respectively. Our goal is that in the new viewing pyramid, the new left-middle and right-middle viewing rays should intersect the sphere/cylinder at the same positions. Therefore, the new left-middle viewing ray, $Ray_{left}^{new}$, is the ray from $P^h$ toward $I_{left}$. The new right-middle viewing ray, $Ray_{right}^{new}$, is the ray from $P^h$ toward $I_{right}$. Finally, we calculate the new "left-horizontal" and "right-horizontal" FOV angles, $fovx_{left}$ and $fovx_{right}$, as the angles between $Dir$ and $Ray_{left}^{new}$ and between $Dir$ and $Ray_{left}^{new}$, respectively. The new vertical FOV angle is calculated as $atan( \frac{(tan(fovx_{left})+tan(fovx_{right}))}{aspect} )$, $aspect$ is the aspect ratio of the original image plane. Note that the new viewing pyramid could be skewed as $fovx_{left}$ and $fovx_{right}$ are not necessarily the same.

\subsubsection{Performance analysis}

\begin{figure}[t]
  \centering
  \includegraphics[width=1\linewidth]{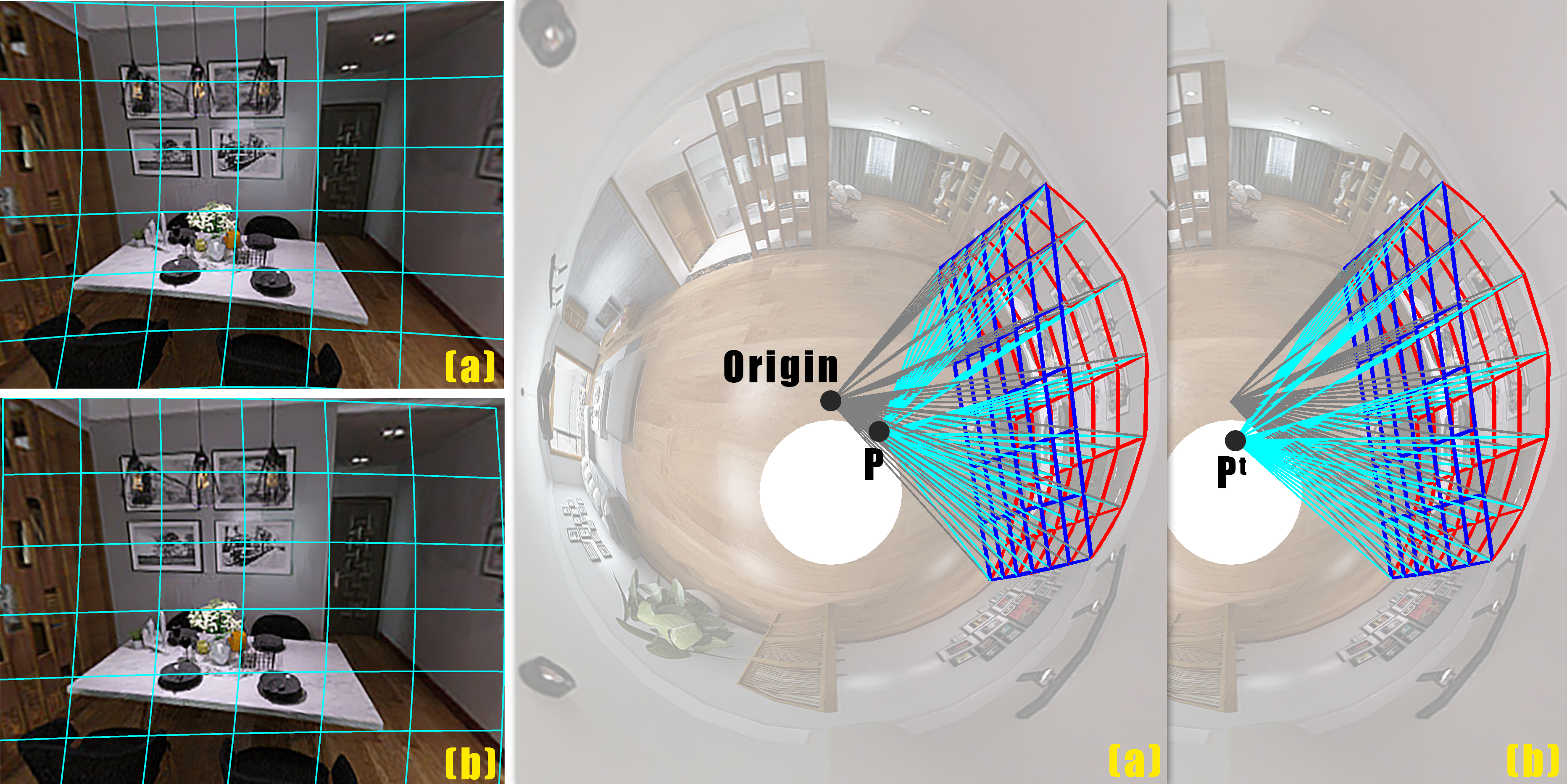}
  \caption{\label{fig:dolly2} Distortion analysis of perspective views of a original camera (a) and the dolly-zoom adjusted version (b). The original camera has a distortion value of $0.00012$ while the adjusted version's amount is just $4.413e-5$. Blue grids come from bilinear interpolations of the four projected corners of the viewing pyramids. Red grids are their projections to the cylinder. Grey lines connect the grid vertices to the origin while cyan lines connect them to the camera positions. In short, the cyan grids in perspective views visualize how regular grids on each camera's image plane (blue) are first projected to the cylinder (red) and then projected back to the respective image planes, causing distortions. We used a coarser grid for visualization with less clutter.} 
\end{figure}

We propose the following scheme to evaluate how the heuristic solution reduces distortions. In short, for a given camera position, looking direction, and FOV angles (a "camera pose" in short), we estimate a subset of the sphere/cylinder that is projected to roughly cover the whole image plane. We subdivide the subset to be a regular grid. We then calculate the curvature of the grid's 2D projection on the image plane as the way to measure the "distortion value" of the camera.

How to estimate such a subset of the sphere/cylinder? We first project the four corners of the view plane onto to the sphere/cylinder. Afterwards, we build a nearly-planar regular grid through bilinear interpolation of the four projected corners. We denote the regular grid as a 2D array of vertices $v[i,j] \in R^3$, $0 \le i \le ROWS, 0 \le j \le COLS$. In practice, the vertices are projected onto the sphere/cylinder in a panoramic image, so the 3D position of the $(i,j)$-th vertex is actually $v'[i,j] = |v[i,j]|$. Note that this means that the grid becomes a discrete curved surface and the 2D projections of its rows and columns of vertices would not be straight unless it is viewed exactly from the origin. Finally, to evaluate the curvature of the grid's 2D projection on the image plane, we sum up a linearity measurement (Equation~\ref{equ:dolly3}) of the consecutive edges of every rows and columns in the grid. See Figure~\ref{fig:dolly2} for an example.

At a particular camera pose, the distortion improvement is the distortion value of the original camera pose minus the distortion value of the dolly zoom-adjusted camera pose. As shown in Figure~\ref{fig:dolly4}, we measure the distortion improvements at every possible camera poses up to rotational and reflective symmetries in a sphere, sampled to avoid clutter. We can see that the improvements vary greatly at different camera poses. For example, distortions are reduced to zero when we "dolly" the camera (i.e., moving forward or backward without changing the looking direction). However, improvements are non-existent if we "truck" the camera (i.e., moving sideways while fixing the looking direction). We leave finding an analytical form to explain the distribution of improvements among different camera poses to future work.

\subsubsection{Optimization-based solution}

\begin{figure}[t]
  \centering
  \includegraphics[width=1\linewidth]{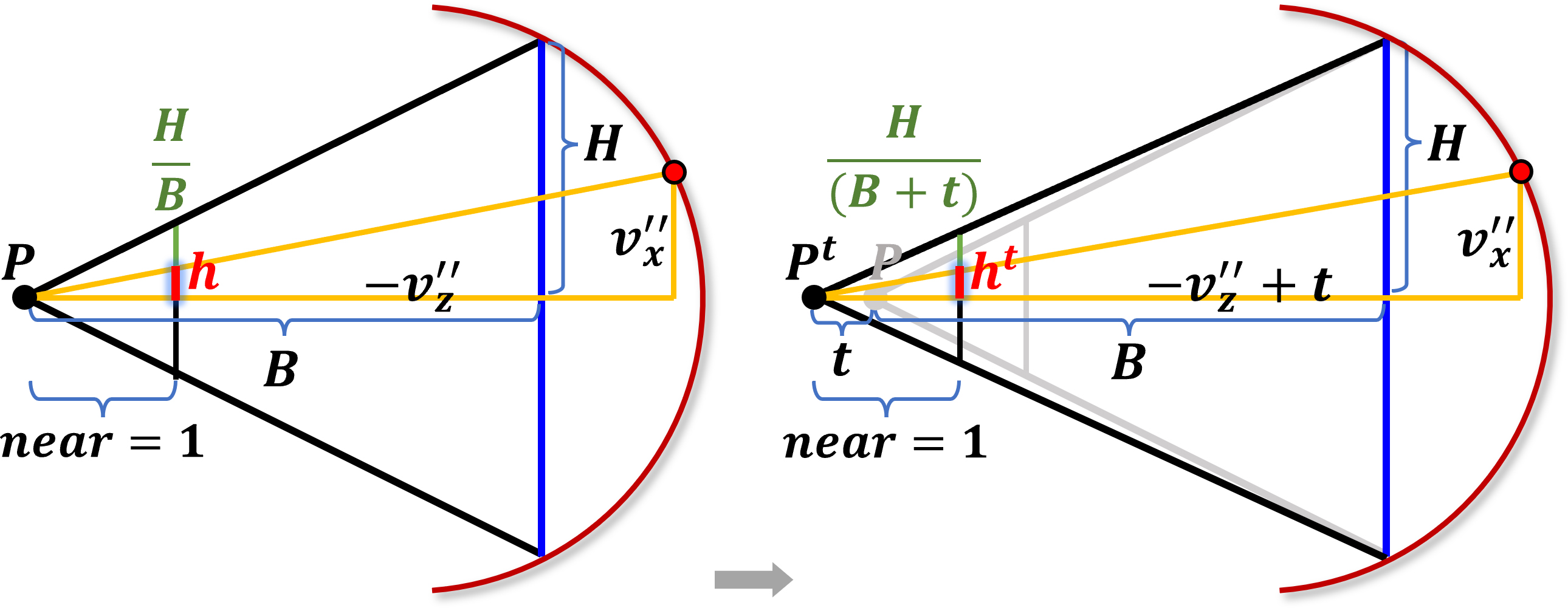}
  \caption{\label{fig:dolly3} The derivation of the 3D-to-2D perspective projection formula (Equation~\ref{equ:dolly1}). We show the derivation of $V_x$ of a point on the cylinder (red dot) and assume $Aspect = 1$. The distance to the view plane, $near$, is 1. Left: an original camera without position offsets (i.e., $t = 0$). $h$ equals $\frac{v''_x}{-v''_z}$ ($v''_z$ is negative). Therefore, $V_x$ equals $\frac{h}{\frac{H}{B}}$ (i.e., we need to normalize $h$ by the half-width of the view plane). Right: the camera is moved along the z-axis by amount $t$. $h$ now equals $\frac{v''_x}{-v''_z + t}$. The new half-width of the view plane becomes $\frac{H}{B+t}$. Therefore, we have $V_x = \frac{v''_x}{\frac{H}{B+t} \cdot (-v''_z + t)} $.} 
\end{figure}

\begin{figure*}[t]
  \centering
  \includegraphics[width=1\linewidth]{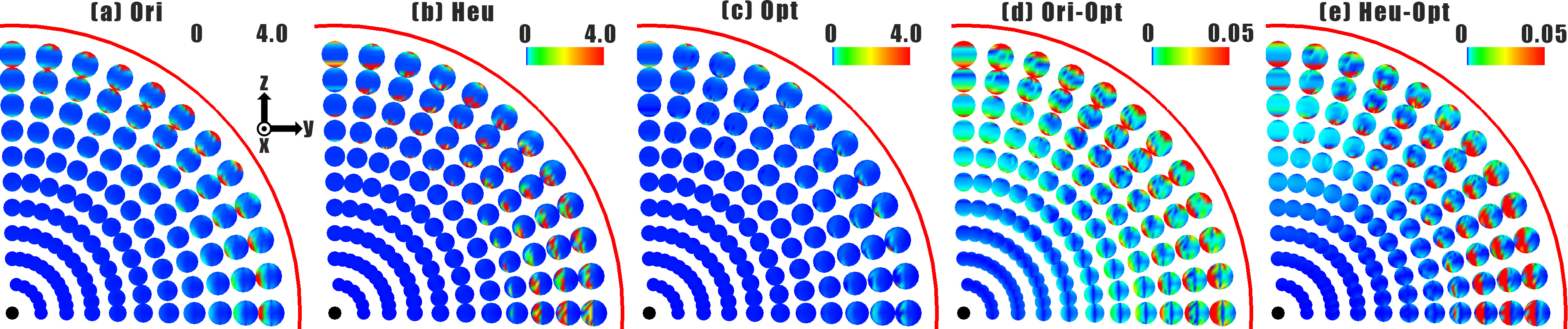}
  \caption{\label{fig:dolly4} Comparing distortions of various camera models. (a) to (c): the distortion values of the original camera model, heuristic solution, and optimized solutions with cylindrical projection. We only show results up to rotational and reflective symmetrices in a sphere. Therefore, we sample the camera positions on a radially sampled first quartile of the Y-Z plane intersect with the unit sphere. At each position, we sample all possible looking directions on the half hemisphere facing the +x direction. The distortion values are converted to colors from blue (smallest) to yellow to red (biggest). For clarity, we cap the upper bound (there exist some very large distortion values so using the full range would result in all blue-ish colors). We can see that optimized solutions clearly have the lowest distortion values among the three. While not clearly shown by the visualizations here, heuristic solutions actually win over the original camera models (see Table~\ref{tab:stat} for statistics). In (d) and (e), we show the distortion improvements of the optimized solutions over the original and the heuristic solutions, respectively.} 
\end{figure*}

The aforementioned evaluation scheme inspired us to formulate the task of finding $t$ as an optimization problem to minimize barrel distortions. In short, we have a single variable $t$ (i.e., offset of the camera position along the looking direction). Given the 3D positions of the grid vertices, $v'[i,j]$, as constants, we first transform them to a new coordinate space that is the same as the view space in OpenGL. That is, the camera position is aligned to the origin, the three axis, $Left$, $Up$, and $Dir$, are aligned to the -x-axis, +y-axis, and -z-axis, respectively, through a rotation and a translation (e.g., by applying OpenGL's MODELVIEW transform matrix). We denote the transformed grid vertex positions as $v''[i,j] \in R^3$. In this manner, we can concisely formulate their 2D projected positions onto the image plane, $V[i,j] \in R^2$, as a function of $t$ as follows:
\begin{equation}
\label{equ:dolly1}
\begin{aligned}
& V[i,j]_x = \frac{v''[i,j]_x}{Aspect \cdot \frac{H}{B + t} \cdot (-v''[i,j]_z + t)}, \\
& V[i,j]_y = \frac{v''[i,j]_y}{\frac{H}{B + t} \cdot (-v''[i,j]_z + t)},
\end{aligned}
\end{equation}
where $B$ denotes the distance of the grid plane to the origin in the new coordinate space. $H$ equals $B \cdot tan(fovy/2)$. $B$ and $H$ can be understood as the base and height of the triangle from the origin to half of the grid plane (see Figure~\ref{fig:dolly3} for an illustration).

The objective function to minimize the sum of the linearity measurement of consecutive edges of every rows and columns in the grid is formulated as follows:
\begin{equation}
\label{equ:dolly2}
\begin{aligned}
& Obj = & \forall_i \; \Sigma_{j=[1,COLS-1]} angle( V[i,j-1], V[i,j], V[i,j+1]) & + \\
& & \forall_j \; \Sigma_{i=[1,ROWS-1]} angle( V[i-1,j], V[i,j], V[i+1,j]) & ,
\end{aligned}
\end{equation}
where $linearity(a,b,c)$, $a, b, c \in R^2$, denotes the linearity measurement of edge $a,b$ and edge $b,c$. We formulate it as:

\begin{equation*}
\label{equ:dolly3}
\begin{aligned}
& linearity(a,b,c) = & pow( (b_x-a_x)(c_y-a_y) - \\
& & (c_x-a_x)(b_y-a_y), 2).
\end{aligned}
\end{equation*}

To sum up, the optimization problem takes the form:
\begin{equation}
\begin{aligned}
& \underset{t}{\text{argmin}}
& & Equation~\ref{equ:dolly2} & \\
& \text{subject to}
& & Equation~\ref{equ:dolly1} & \forall \; V[i,j] \in grid.
\end{aligned}
\end{equation}
After $t$ is solved, the new camera position and FOV angles are derived accordingly.

In Figure~\ref{fig:dolly4}, we compare the distortion improvements of the original camera, heuristic solutions, and the optimized solutions at different camera poses. We find that the optimized solutions improves upon both the original and the heuristic solutions. In our comparisons, we used 10 for $ROWS$ and $COLS$. We experimented with other grid resolutions and the results are similar. We conclude that using the optimization approach led to better results at a small computational cost.


\section{Results and comparisons}
\label{sec:result}

\begin{figure}[t]
\centering
  
\begin{minipage}{.326\linewidth}
\includegraphics[width=\textwidth]{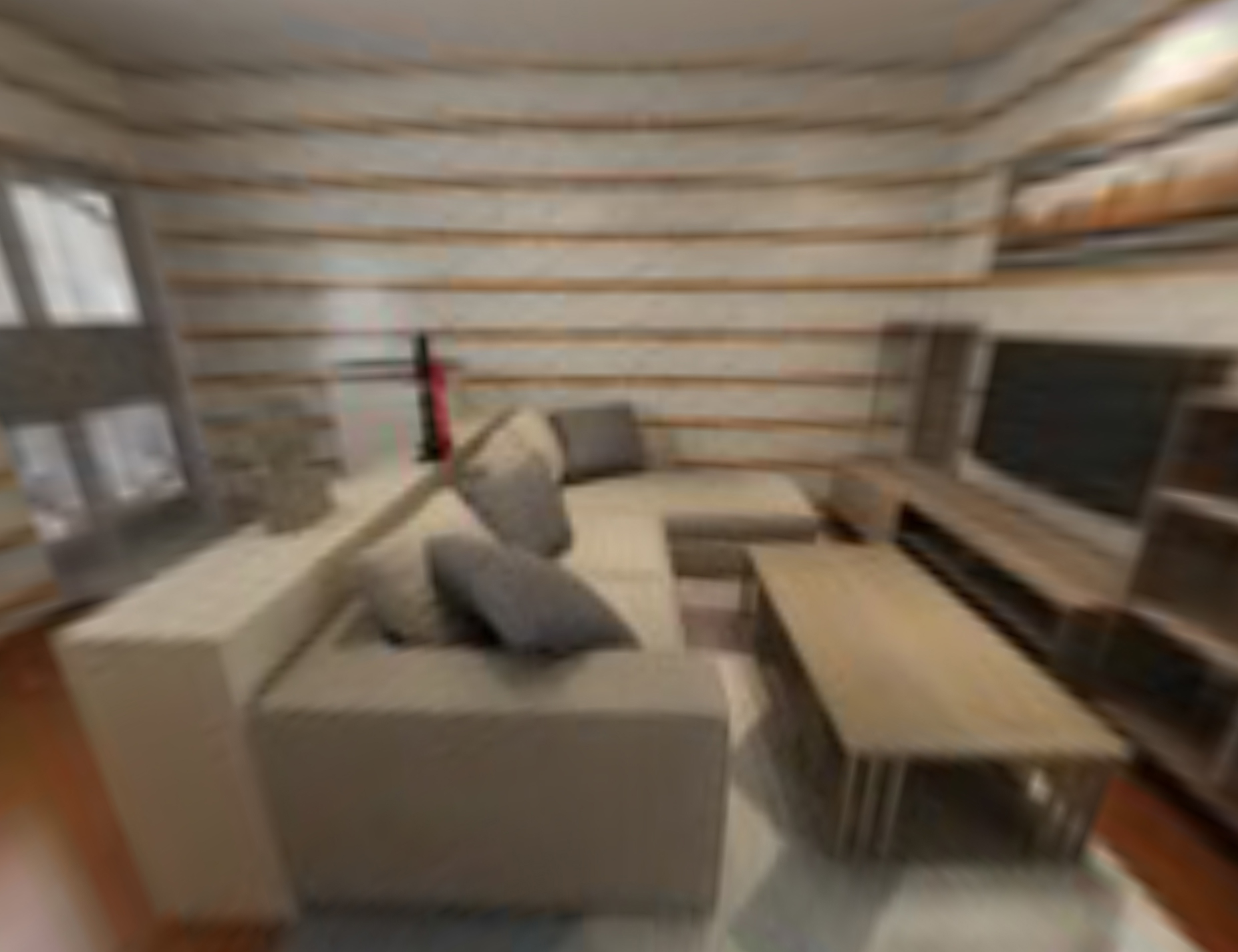}
\end{minipage}
\begin{minipage}{.326\linewidth}
\includegraphics[width=\textwidth]{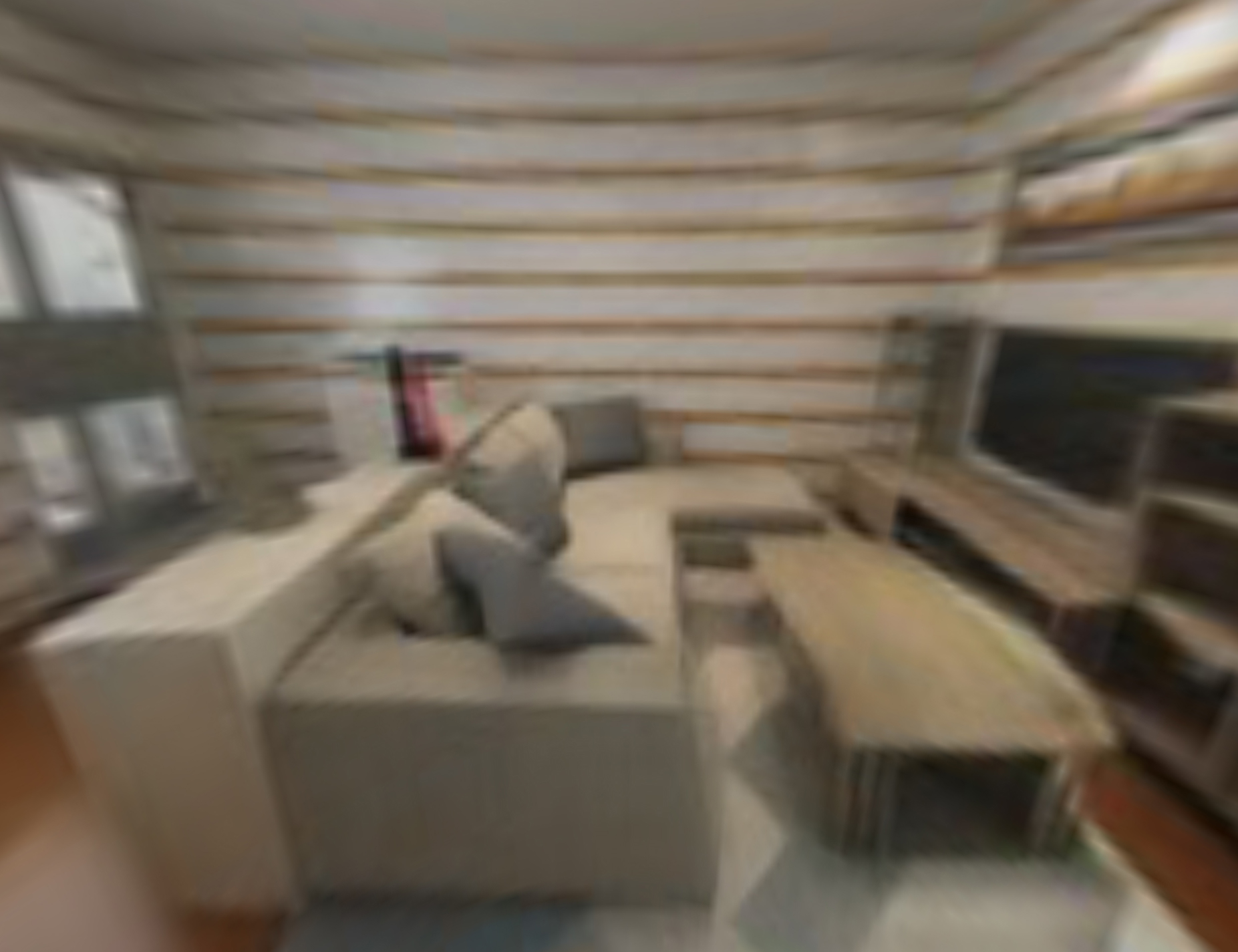}
\end{minipage}
\begin{minipage}{.326\linewidth}
\includegraphics[width=\textwidth]{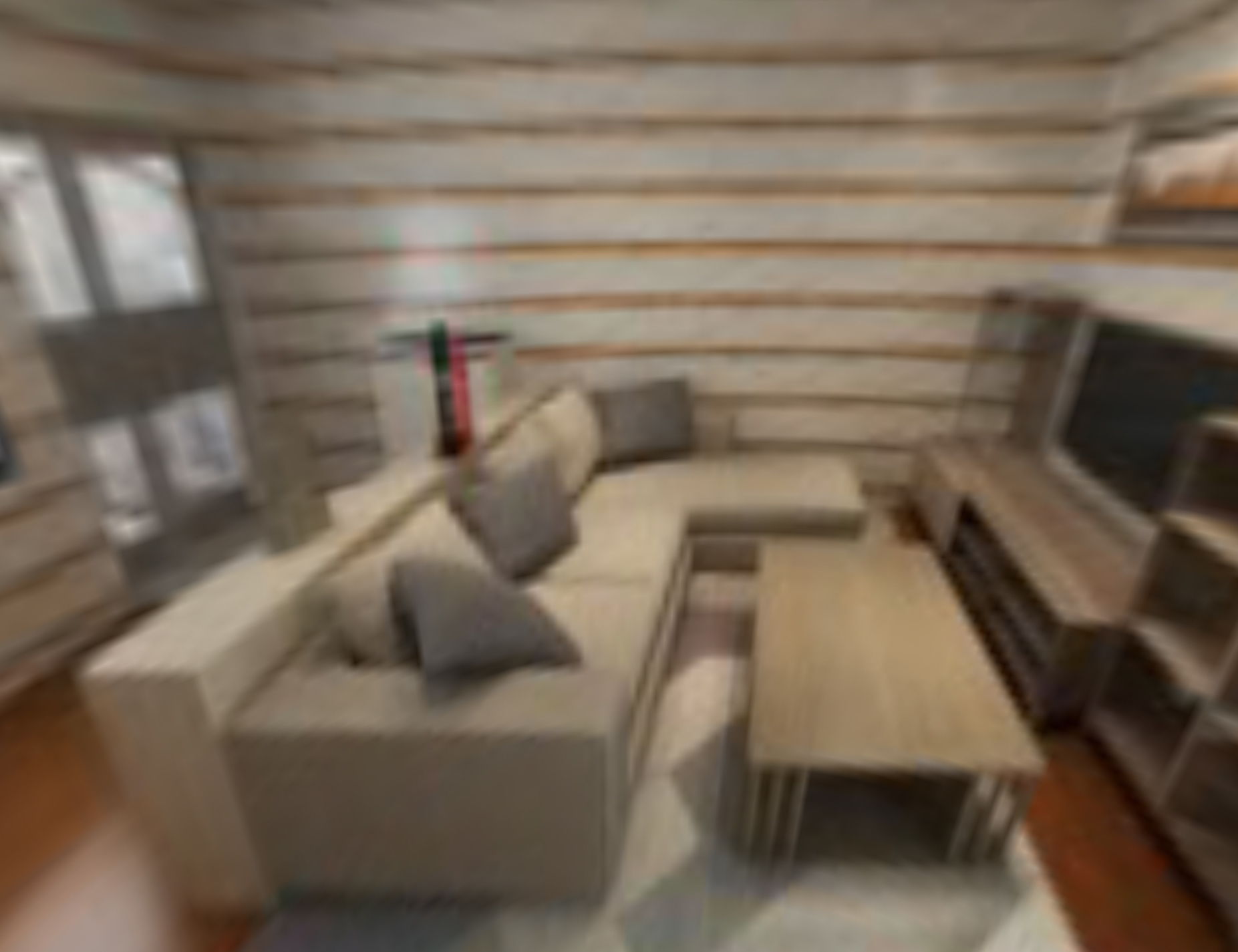}
\end{minipage}

\begin{minipage}{.326\linewidth}
\includegraphics[width=\textwidth]{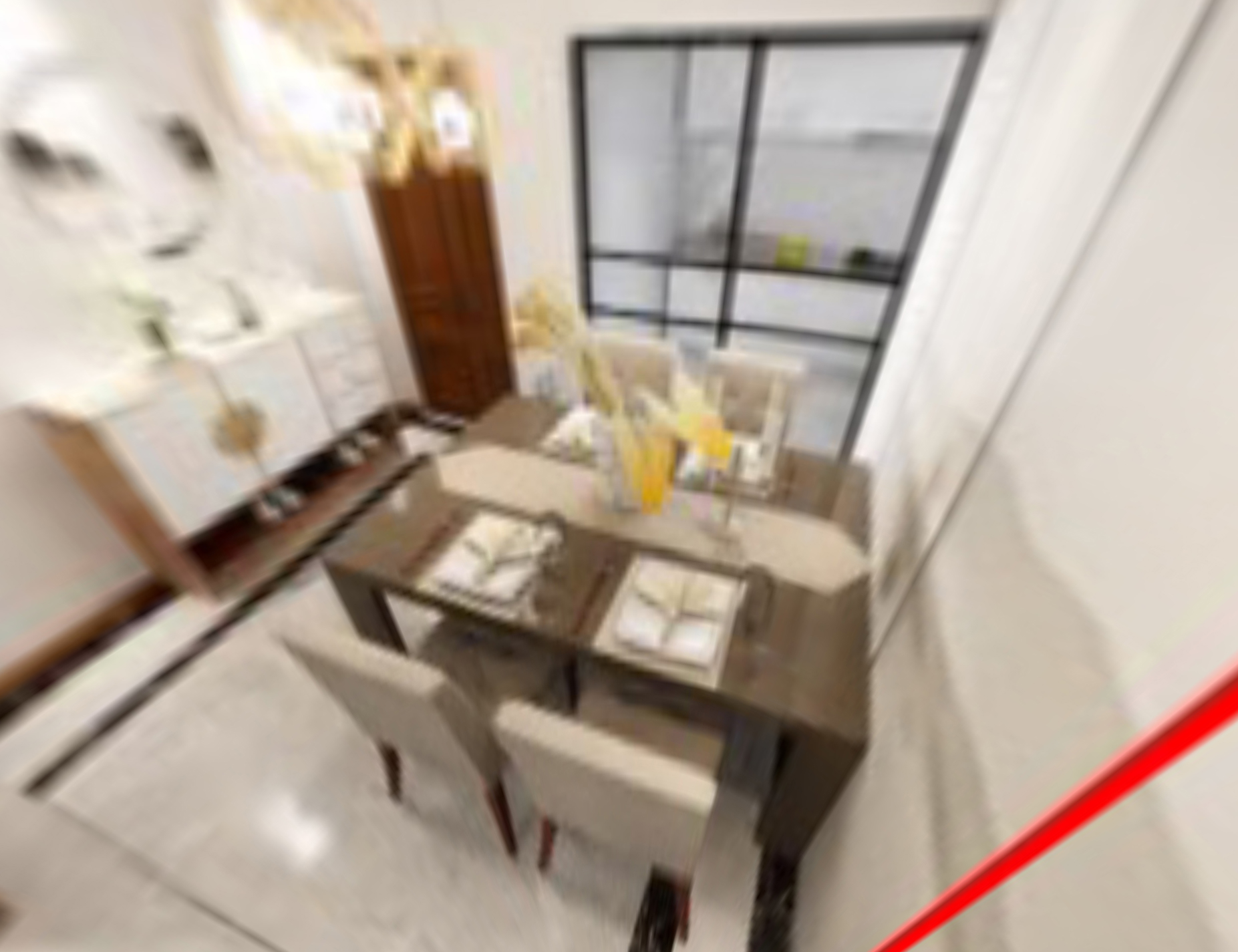}
\end{minipage}
\begin{minipage}{.326\linewidth}
\includegraphics[width=\textwidth]{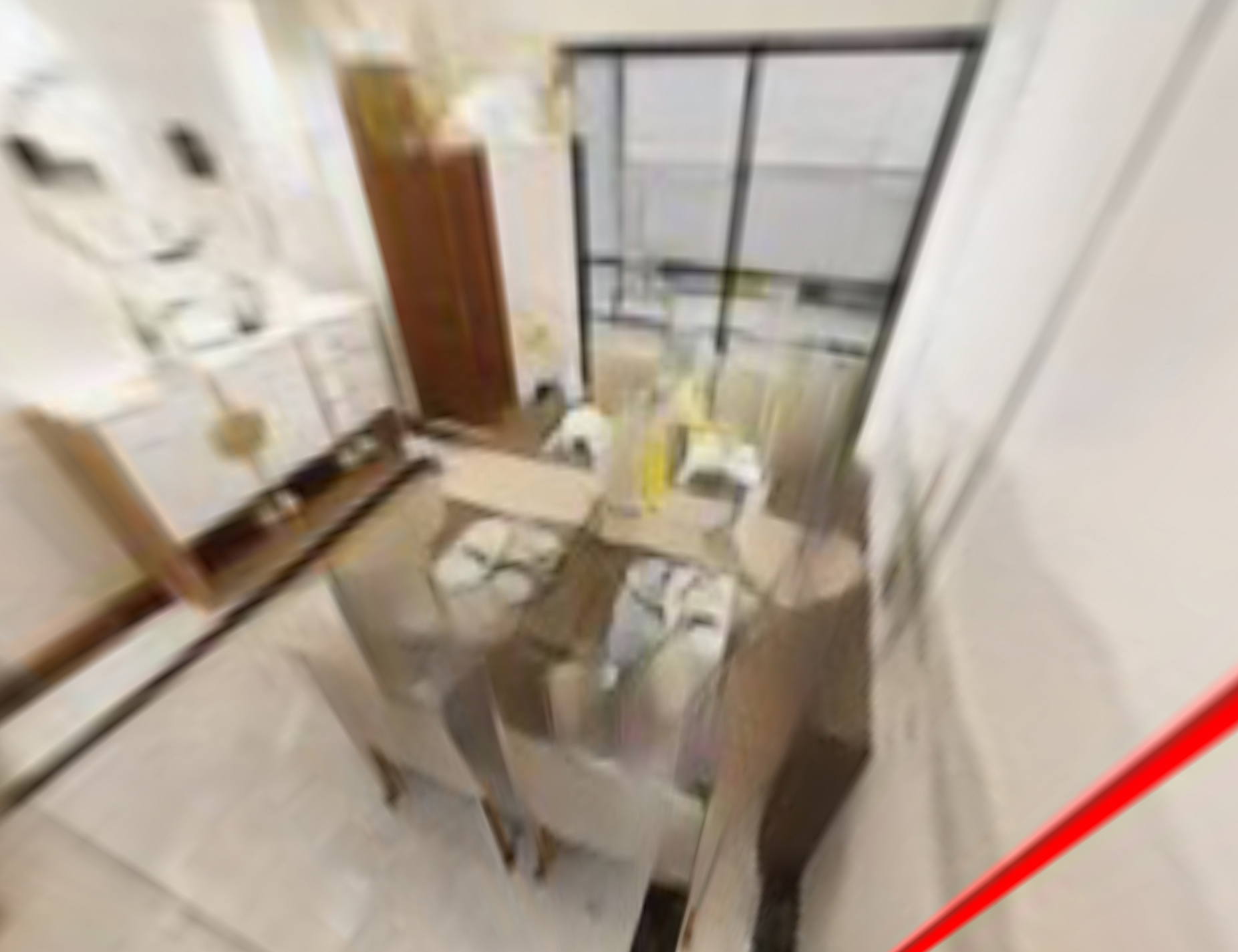}
\end{minipage}
\begin{minipage}{.326\linewidth}
\includegraphics[width=\textwidth]{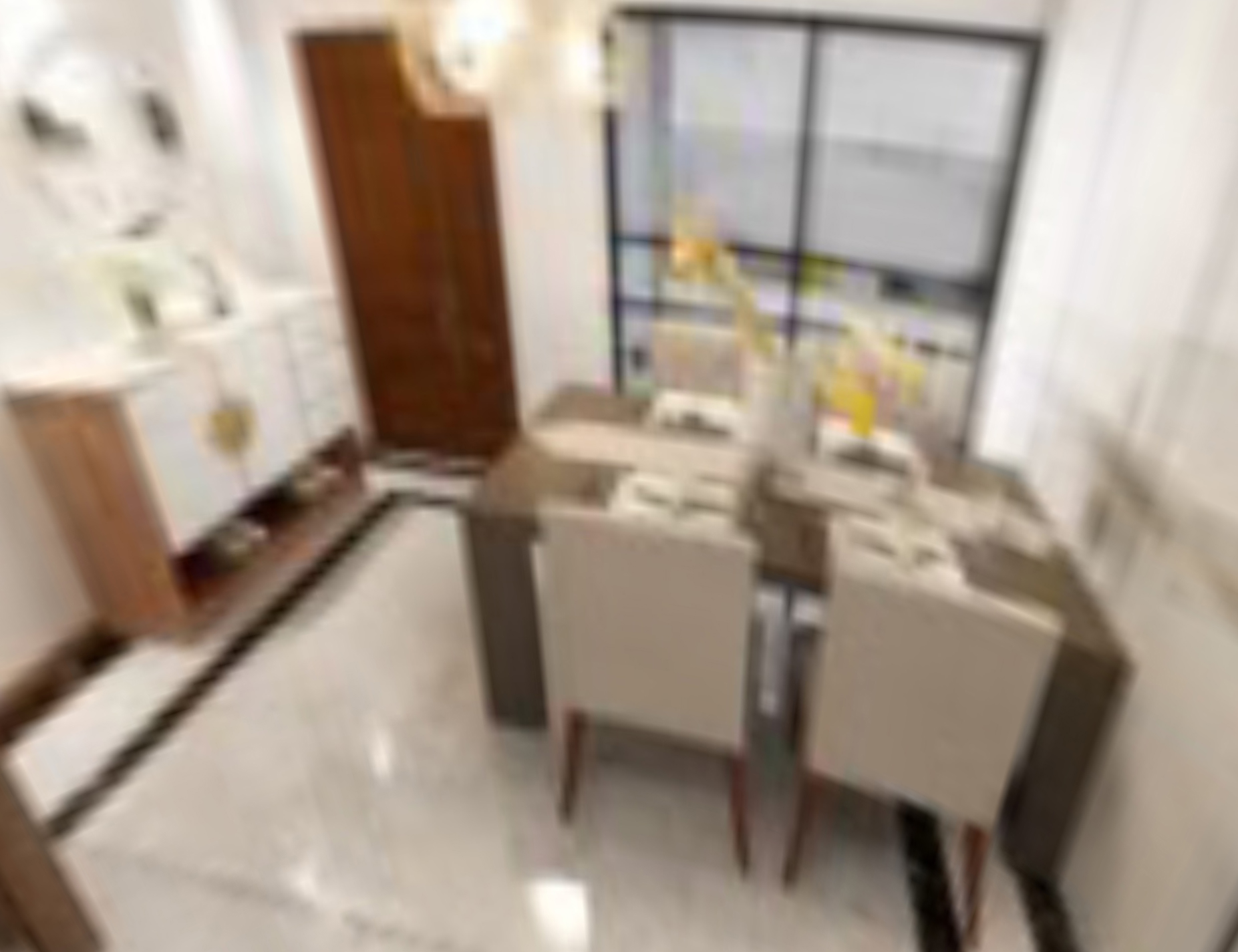}
\end{minipage}

\begin{minipage}{.326\linewidth}
\includegraphics[width=\textwidth]{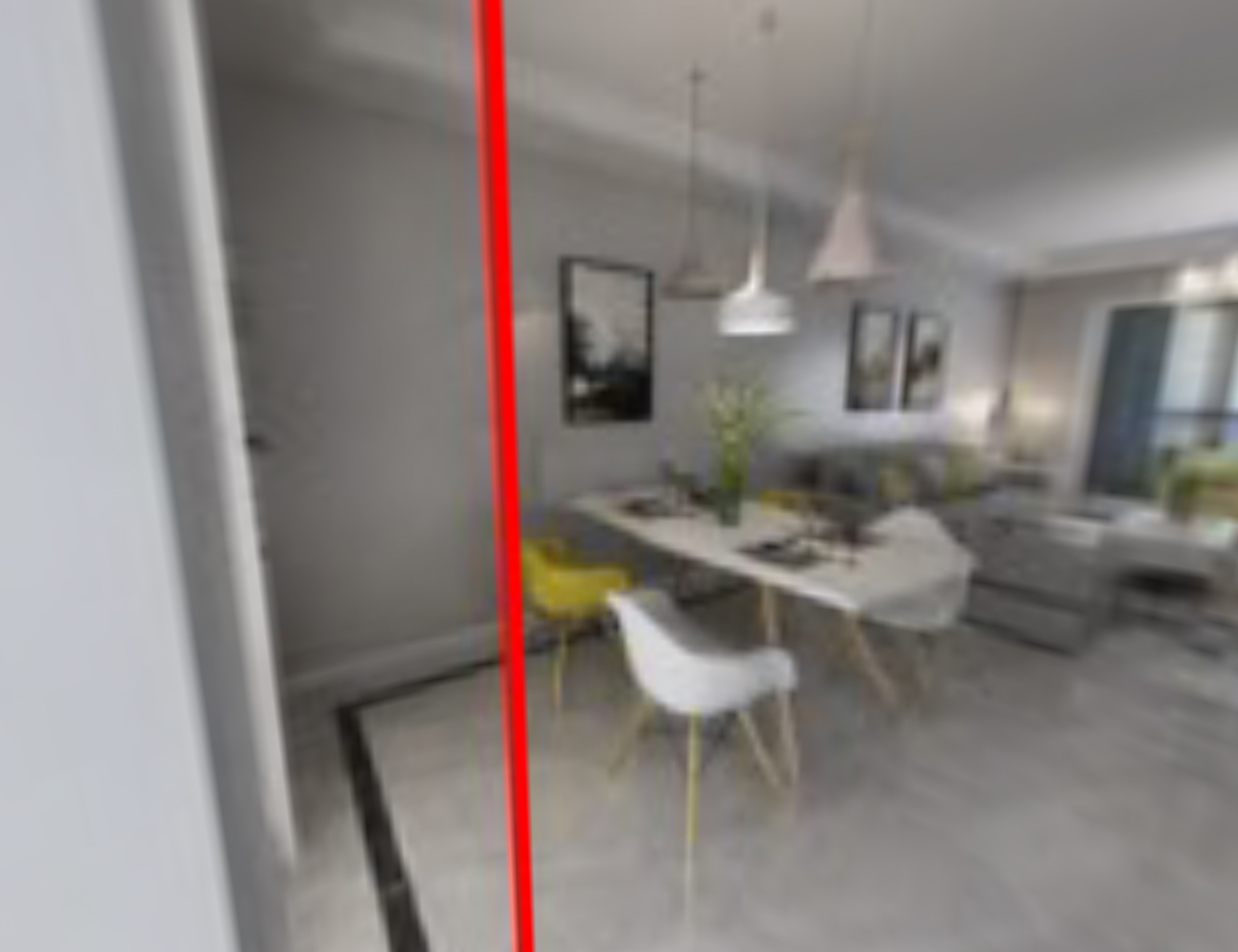}
\end{minipage}
\begin{minipage}{.326\linewidth}
\includegraphics[width=\textwidth]{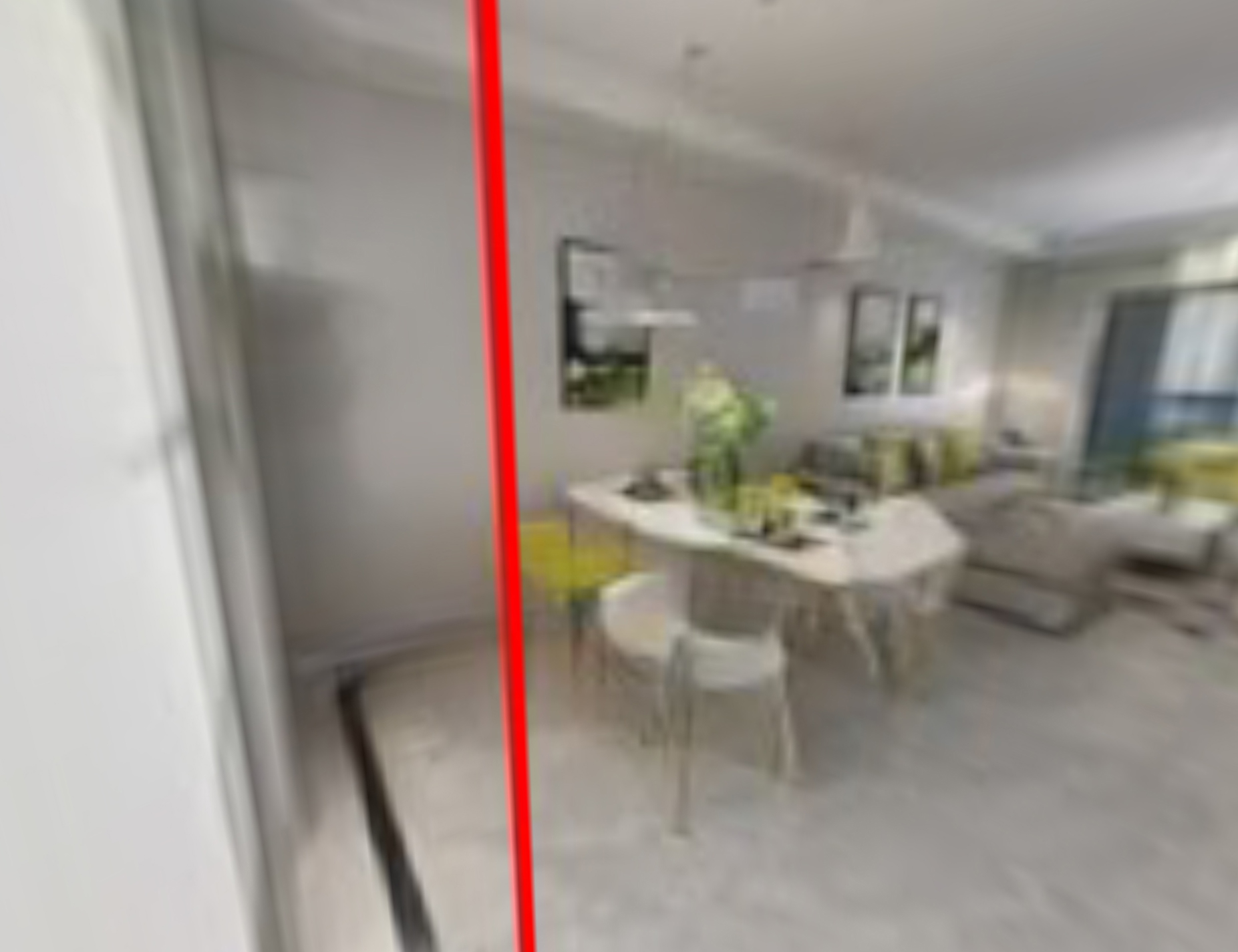}
\end{minipage}
\begin{minipage}{.326\linewidth}
\includegraphics[width=\textwidth]{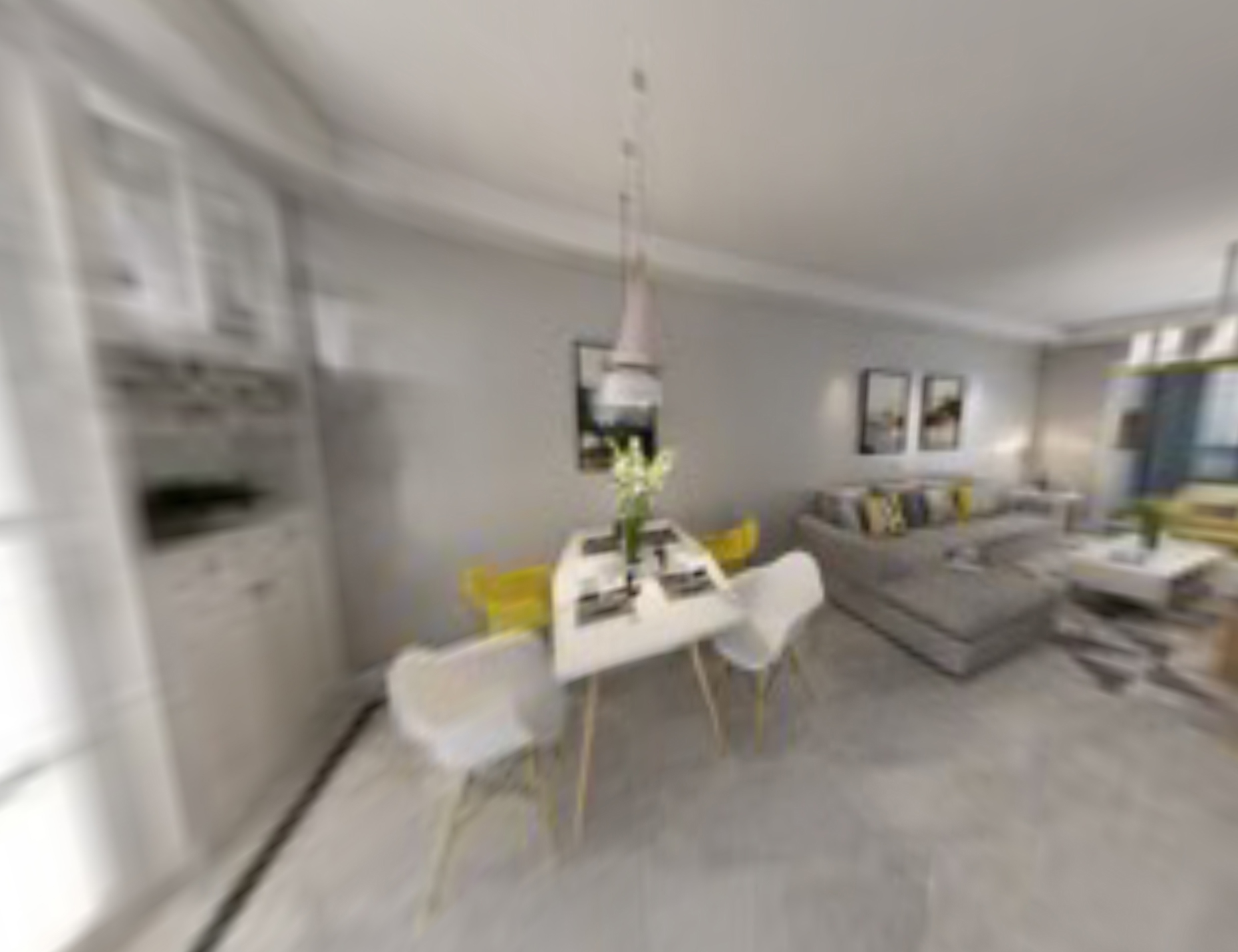}
\end{minipage}
\caption{Our attempts to recreate the same novel view synthesis results in Figure 5 in~\cite{xu2021layout}. For each row, left: a perspectives view of a ground truth "target" panorama. Middle: the same perspective view of the synthesized panorama generated by their method using a different "source" panorama and the camera pose of the "target" panorama. Right: our result with our best effort to find the corresponding camera pose. Our results aren't as geometrically correct as theirs but are free of blurs and ghosts. The red markers can be ignored.}
\label{fig:xuu}
\end{figure}

We first compare our method to~\cite{xu2021layout}. As shown in Figure~\ref{fig:xuu}, we attempt to create the same novel view synthesis results as done in their paper. In general, we find their results to be quite good in getting the overall 3D layouts correct. However, two shortcomings are: 1) the occasional blurs and ghosts, and 2) the resolution of their predicted panoramas is only 512x256, too low for drawing perspective views. In comparison, our results are not as geometrically accurate as theirs, but are of higher resolutions (since the source panoramas tend to have higher resolutions than the predicted panoramas), and largely free of any visual artifacts.

\begin{figure*}
    \centering
\begin{minipage}{.15\textwidth}
\includegraphics[width=\textwidth]{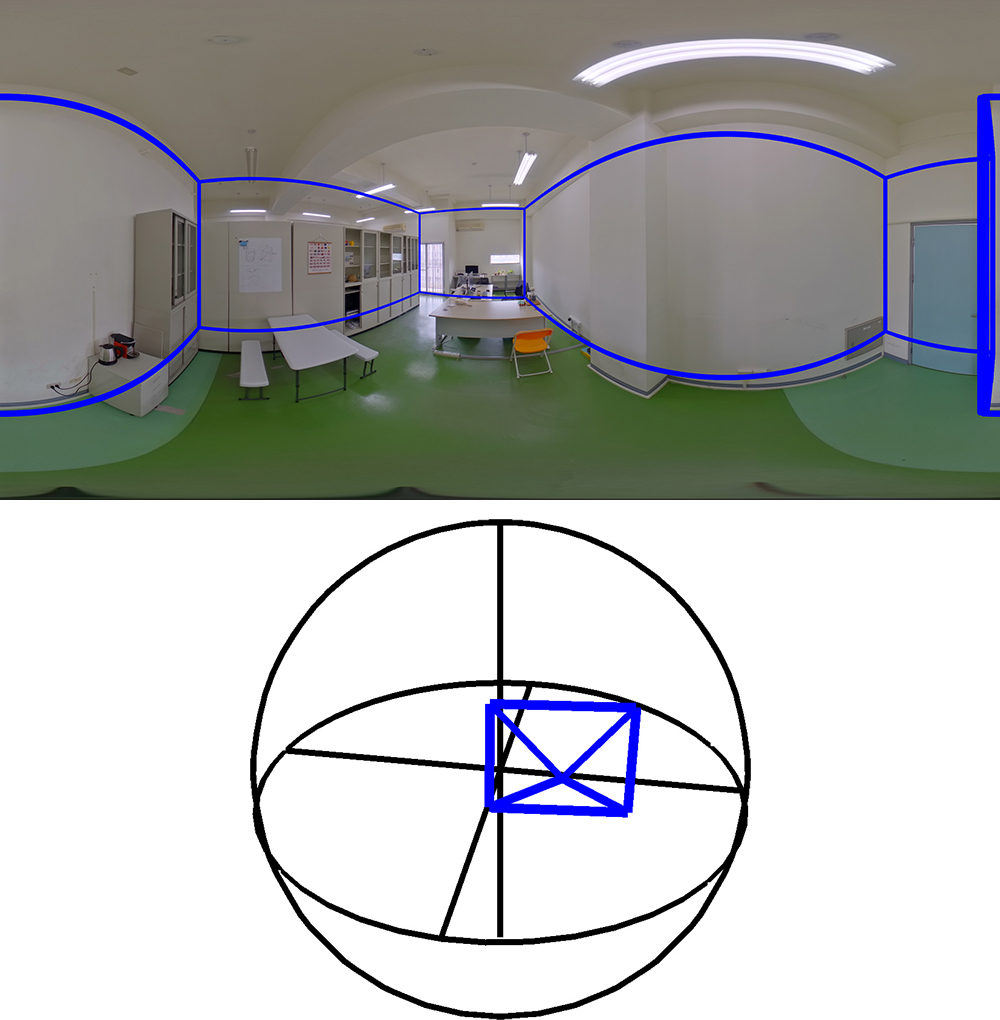}
\end{minipage}
\begin{minipage}{.208\textwidth}
\includegraphics[width=\textwidth]{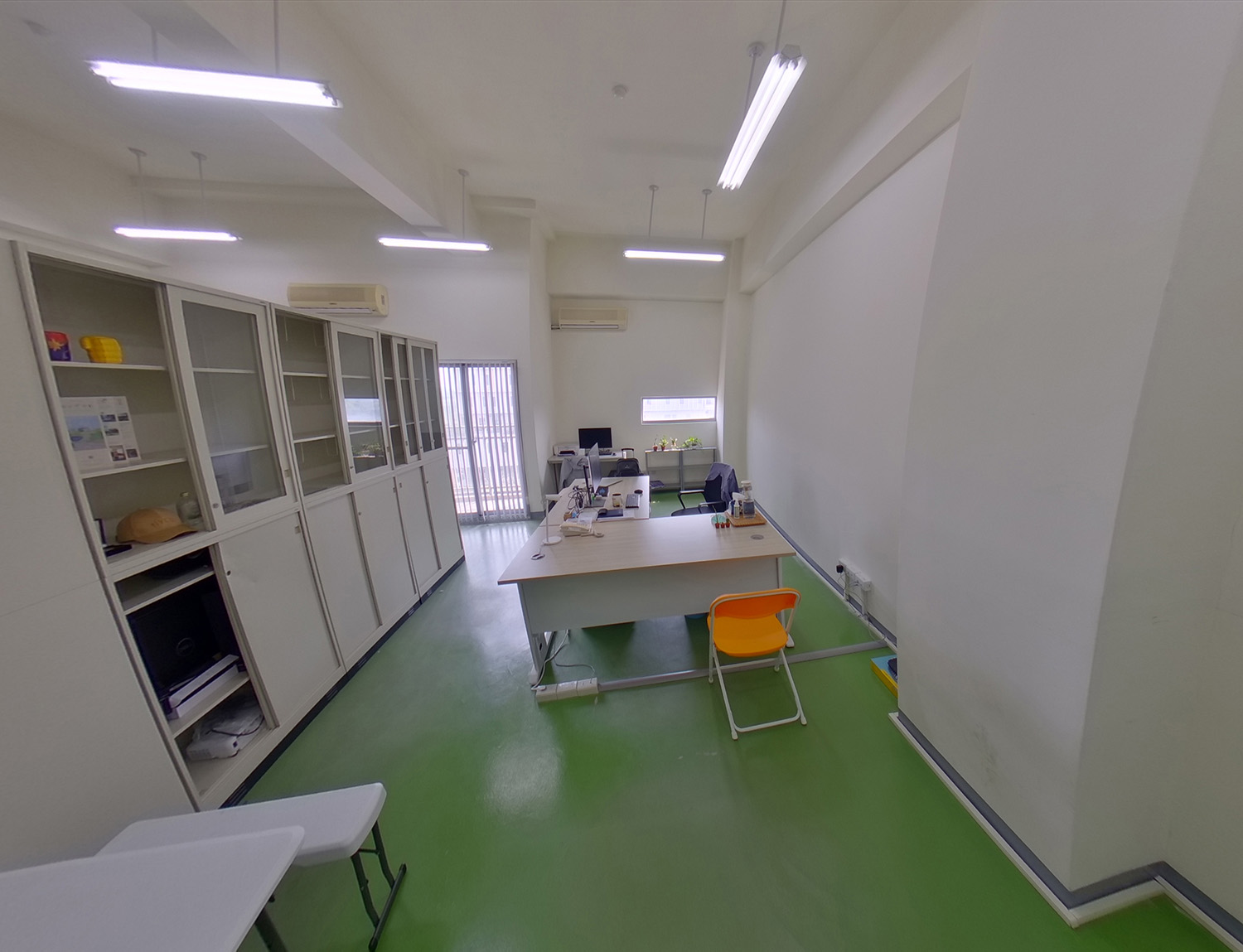}
\end{minipage}
\begin{minipage}{.208\textwidth}
\includegraphics[width=\textwidth]{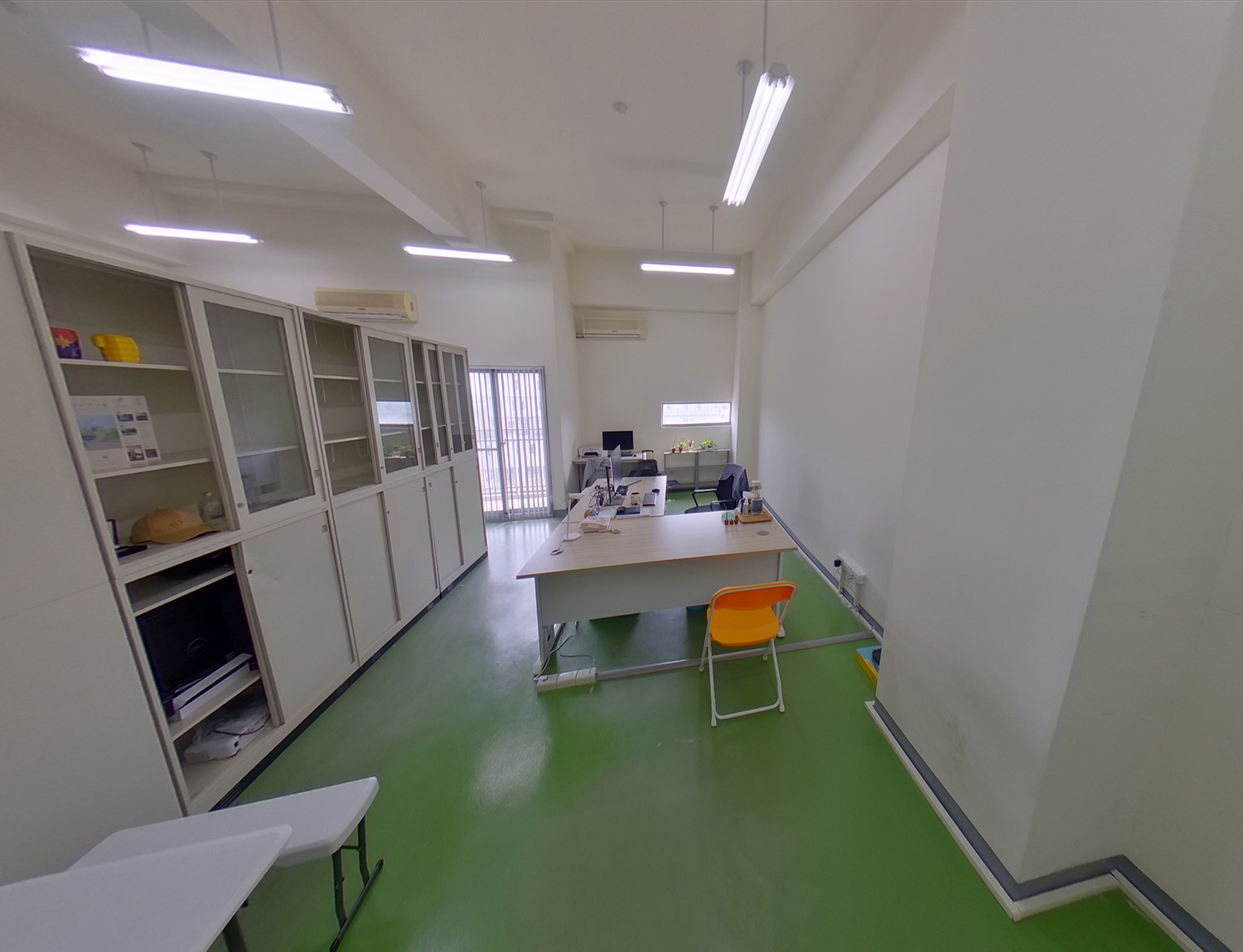}
\end{minipage}
\begin{minipage}{.208\textwidth}
\includegraphics[width=\textwidth]{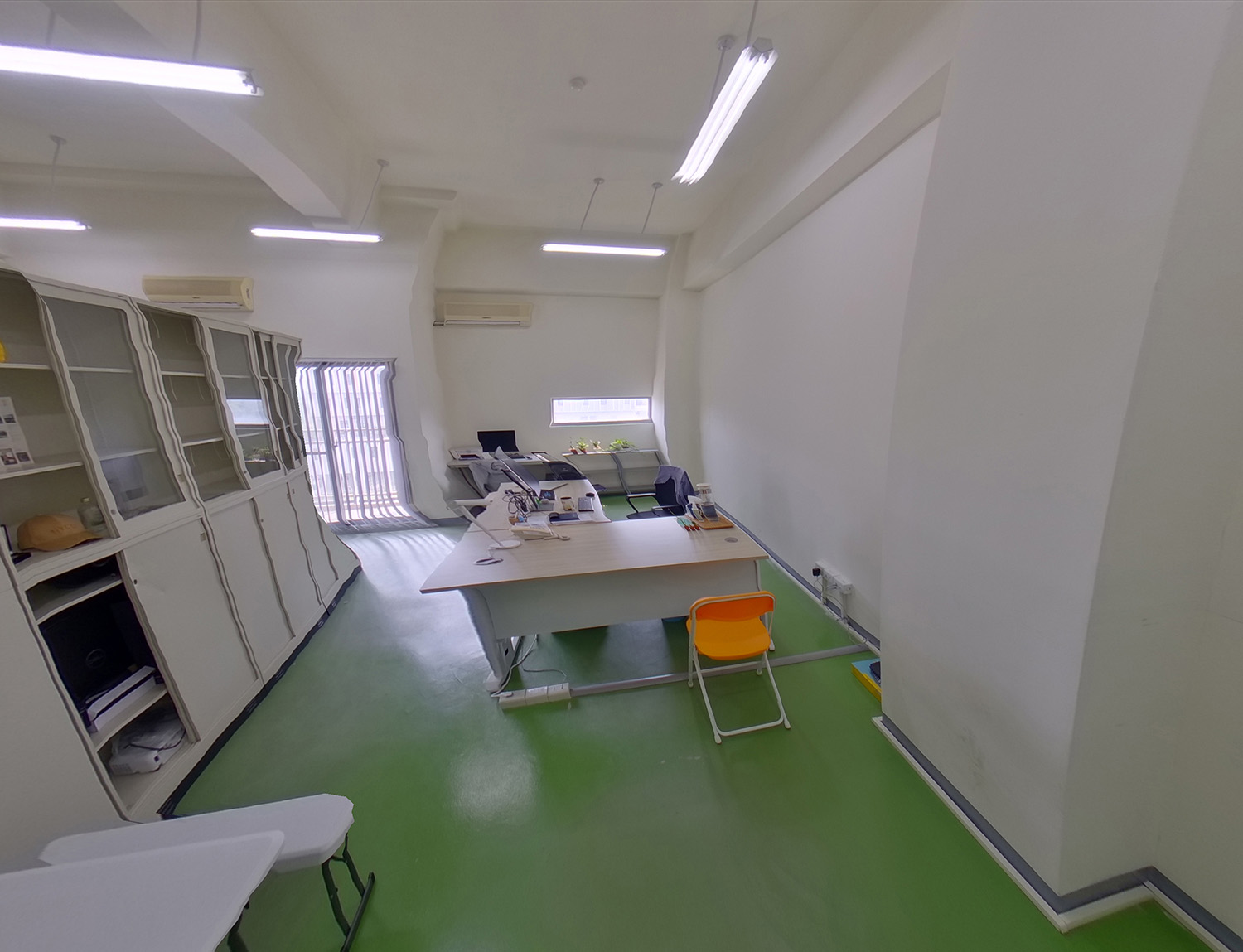}
\end{minipage}
\begin{minipage}{.208\textwidth}
\includegraphics[width=\textwidth]{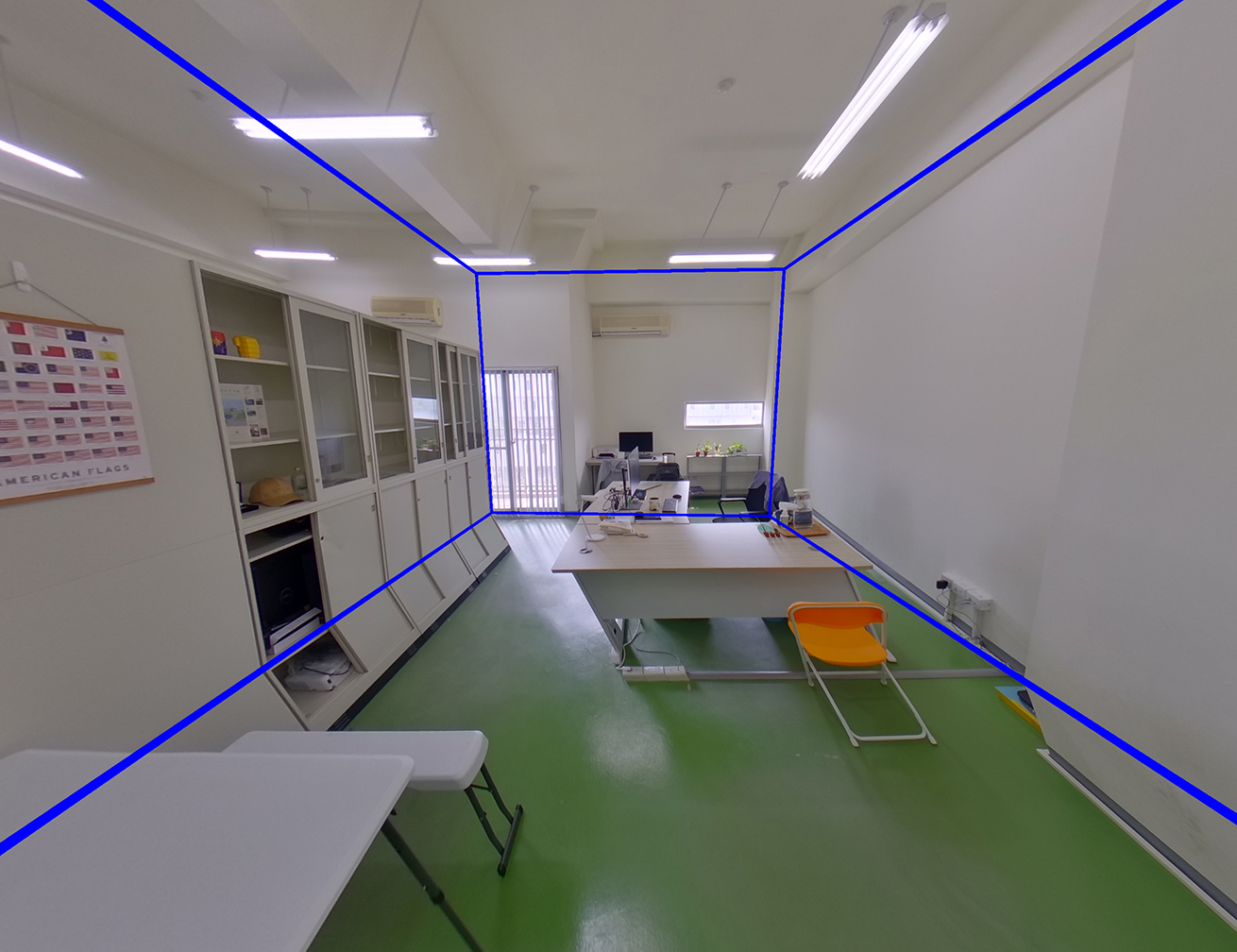}
\end{minipage}

\begin{minipage}{.15\textwidth}
\includegraphics[width=\textwidth]{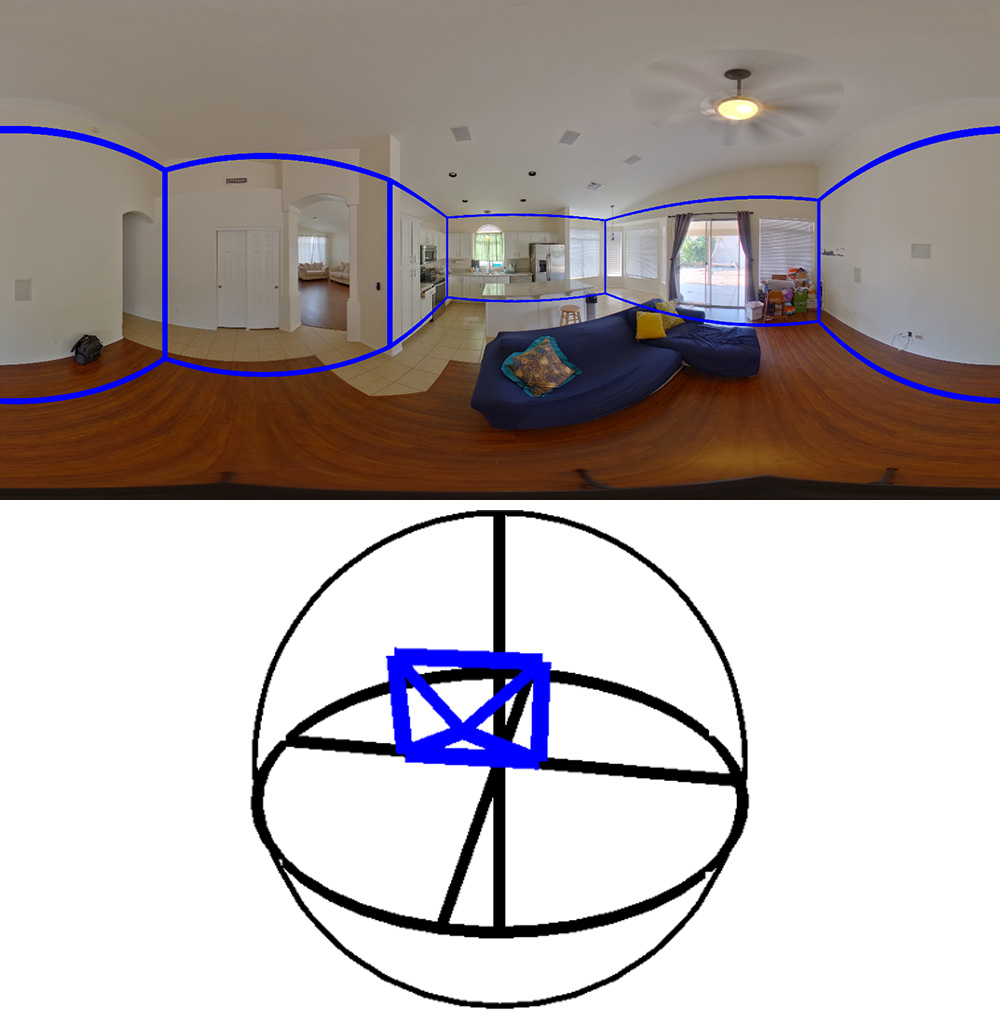}
\end{minipage}
\begin{minipage}{.208\textwidth}
\includegraphics[width=\textwidth]{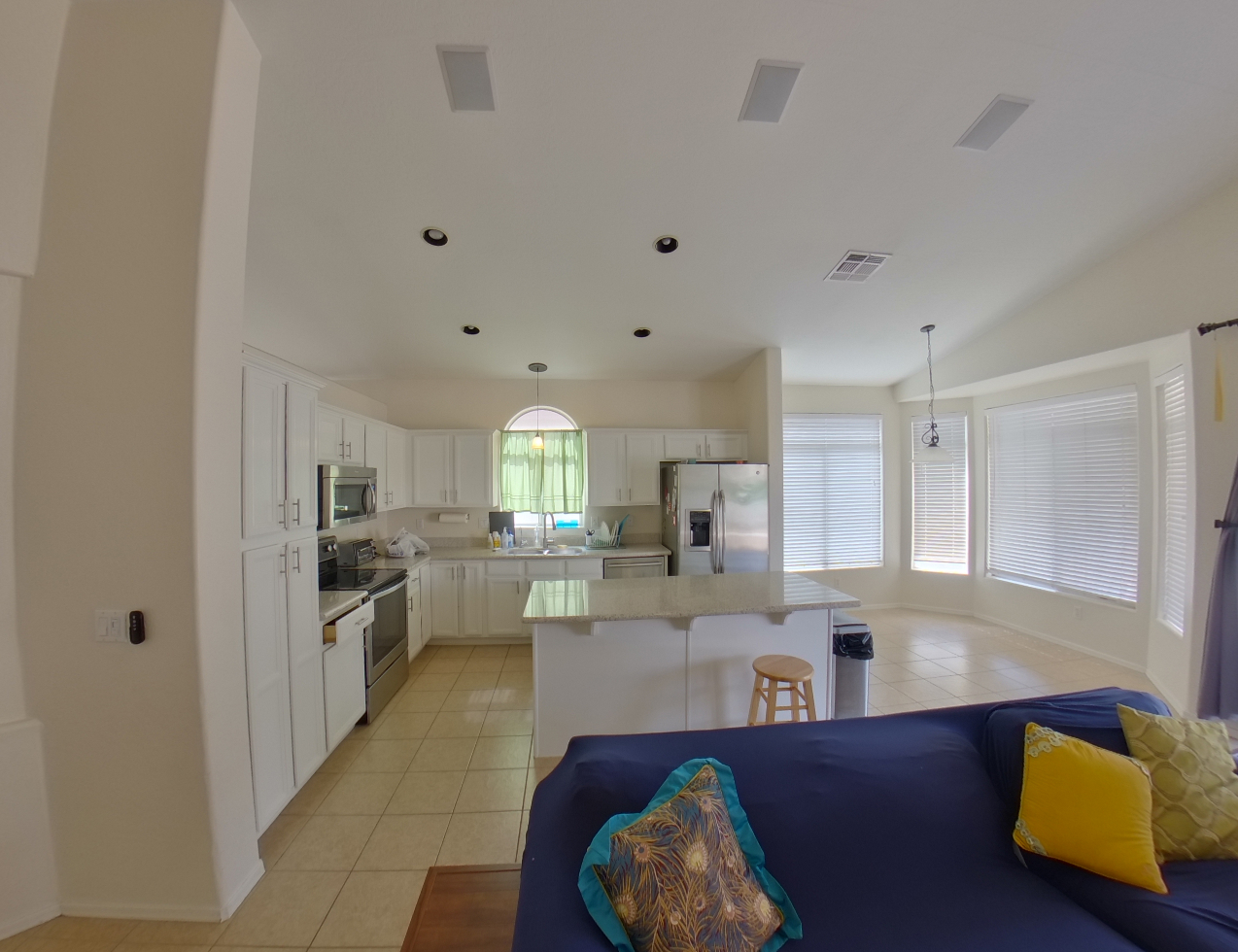}
\end{minipage}
\begin{minipage}{.208\textwidth}
\includegraphics[width=\textwidth]{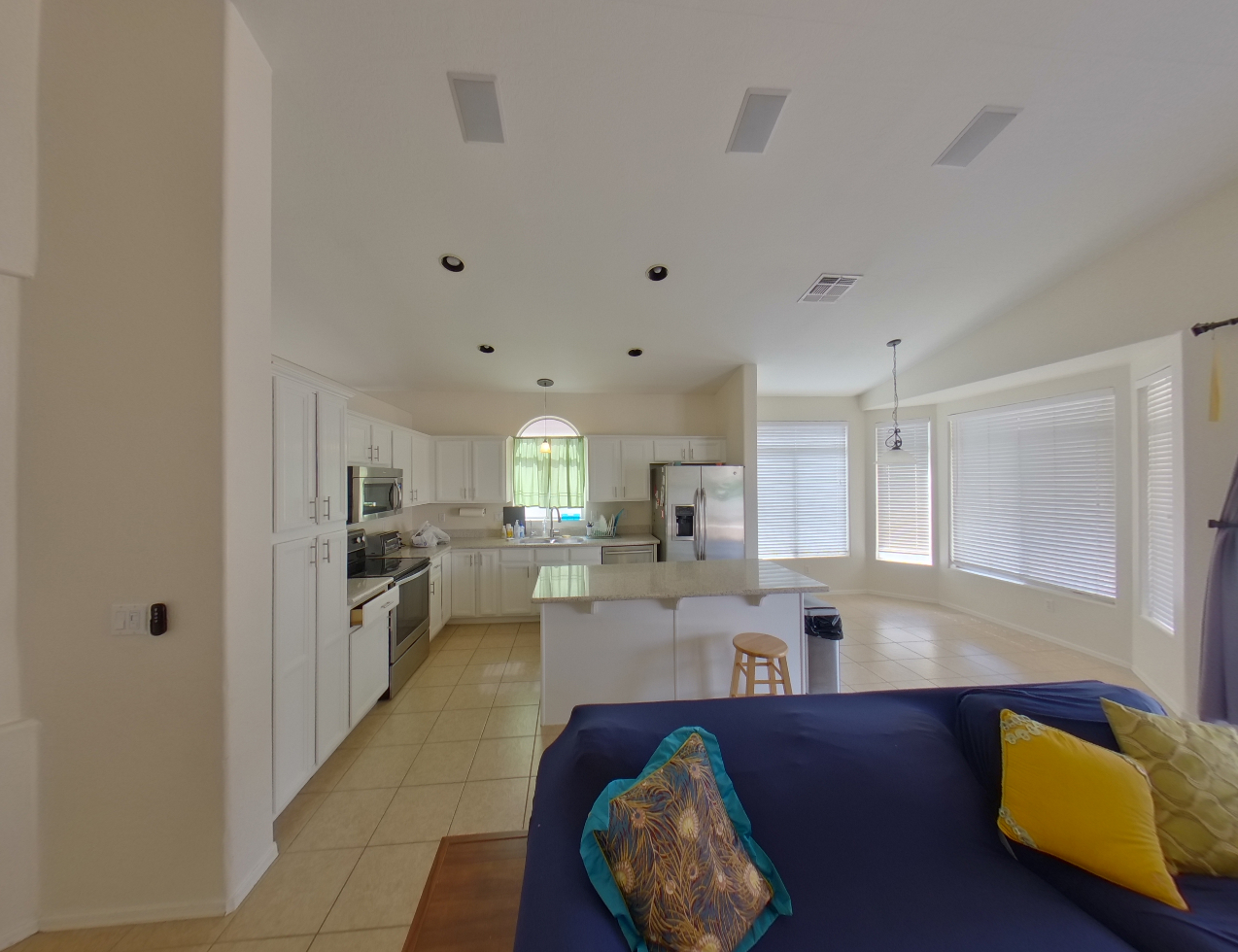}
\end{minipage}
\begin{minipage}{.208\textwidth}
\includegraphics[width=\textwidth]{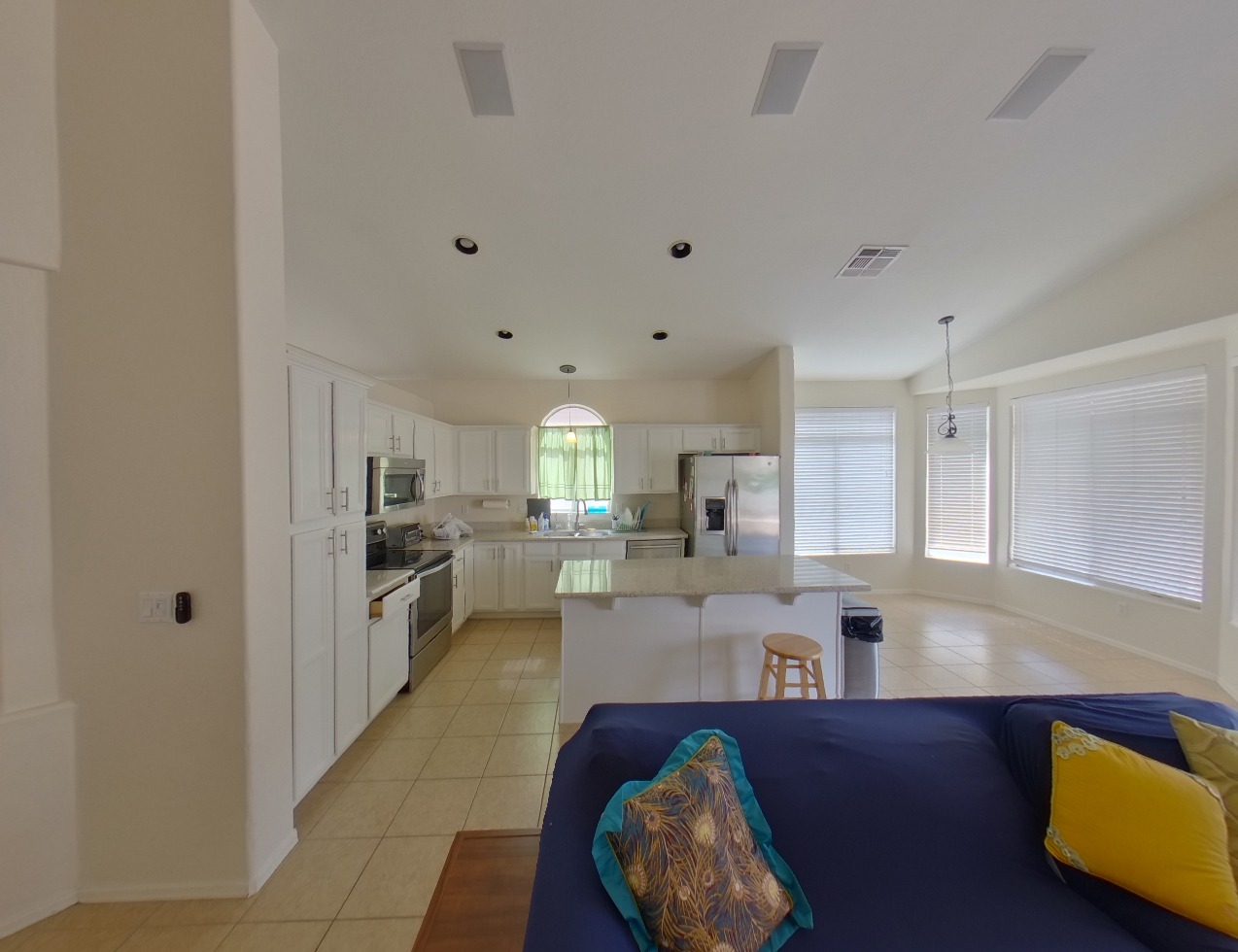}
\end{minipage}
\begin{minipage}{.208\textwidth}
\includegraphics[width=\textwidth]{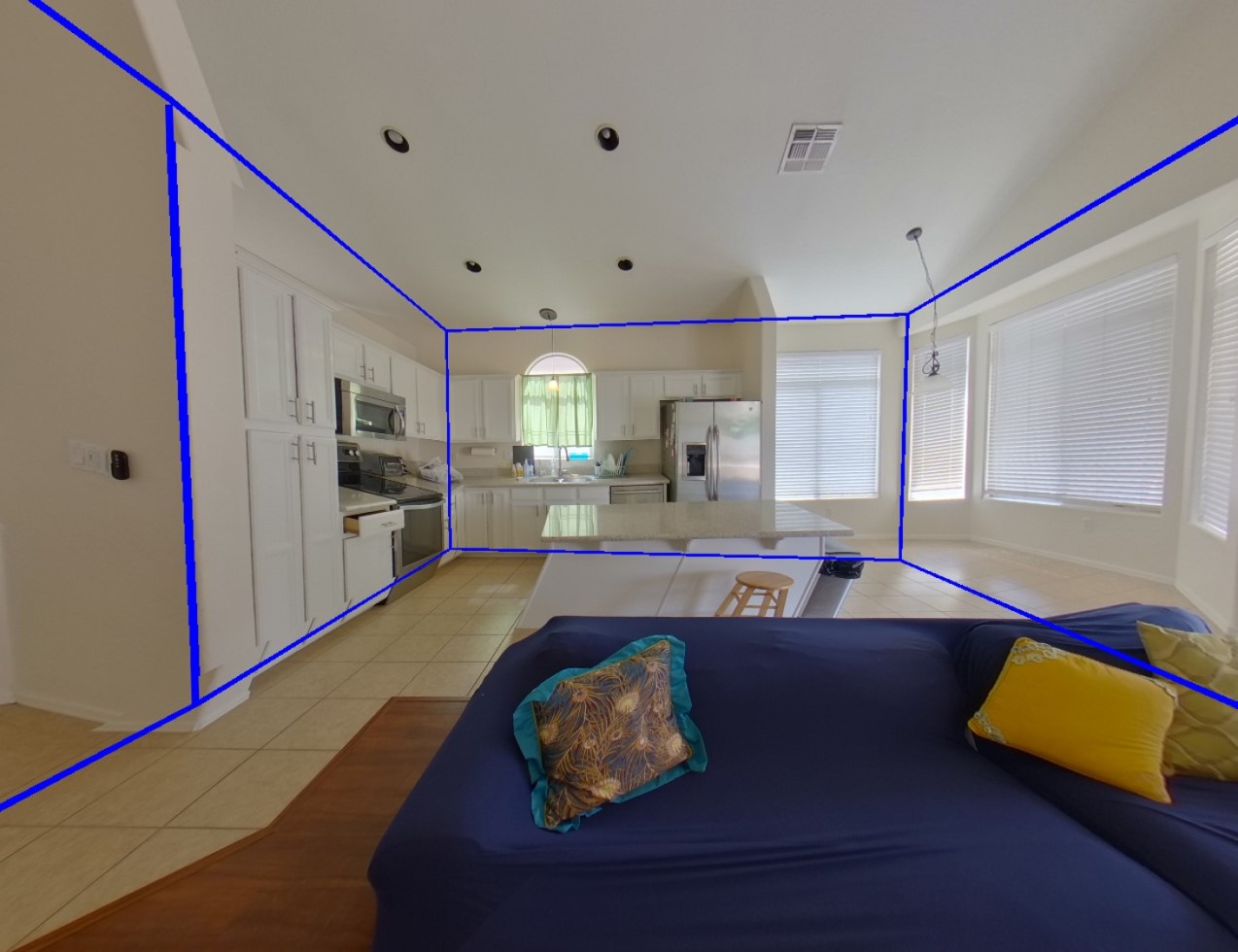}
\end{minipage}

\begin{minipage}{.15\textwidth}
\includegraphics[width=\textwidth]{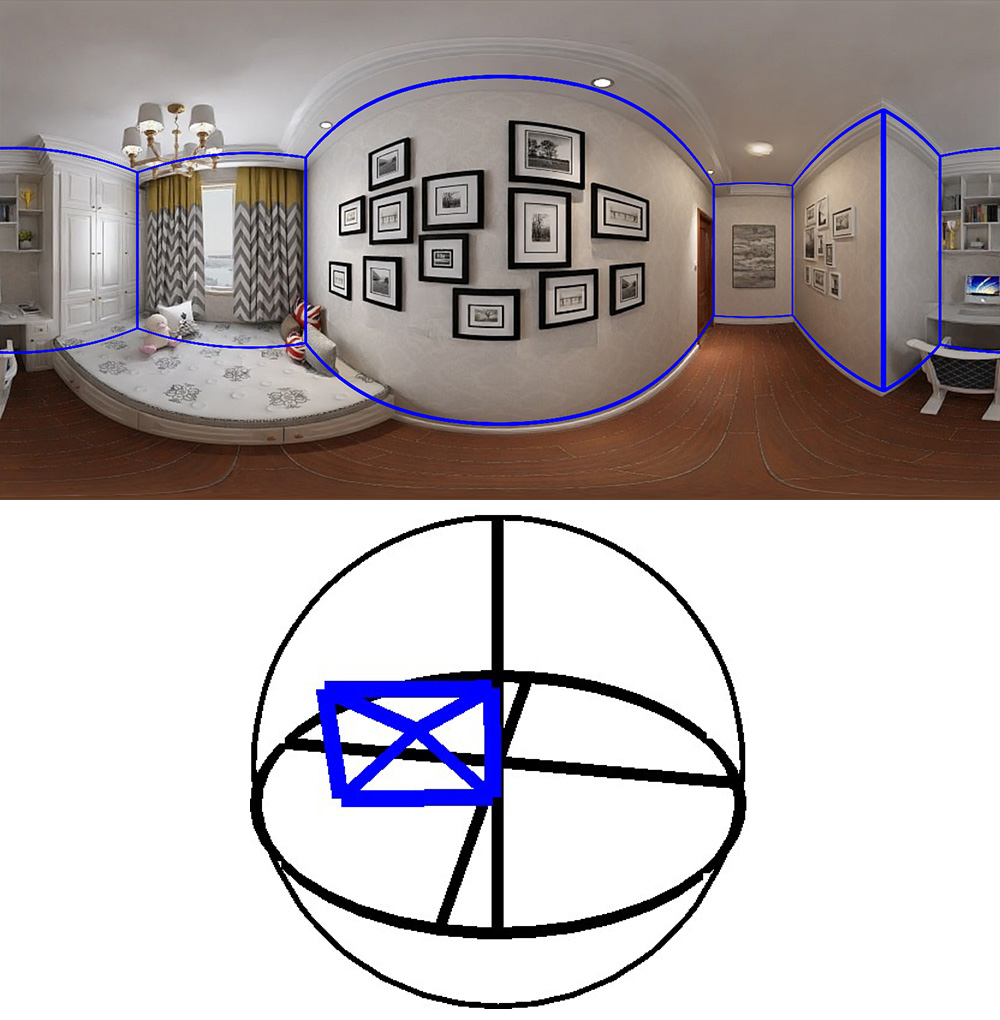}
\end{minipage}
\begin{minipage}{.208\textwidth}
\includegraphics[width=\textwidth]{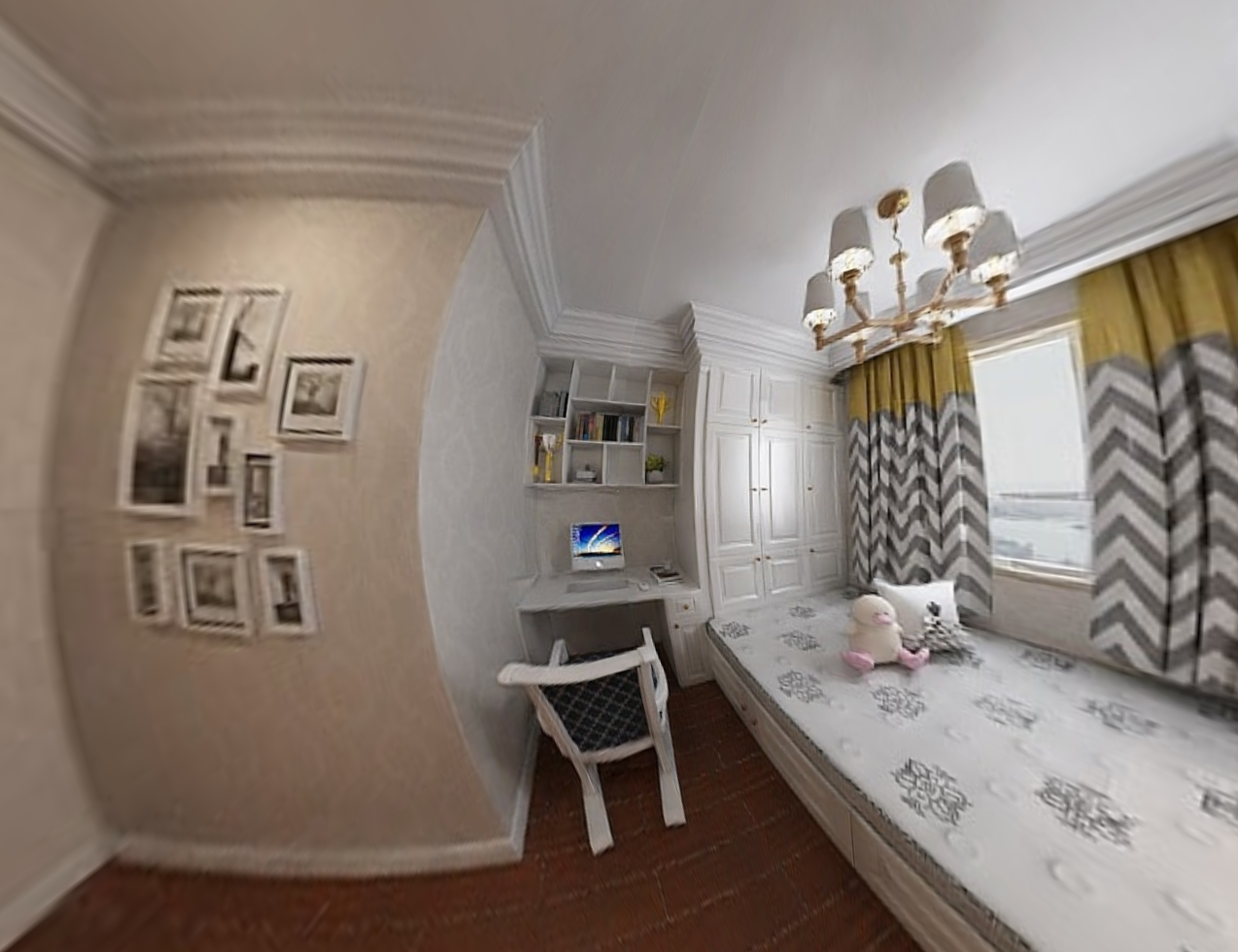}
\end{minipage}
\begin{minipage}{.208\textwidth}
\includegraphics[width=\textwidth]{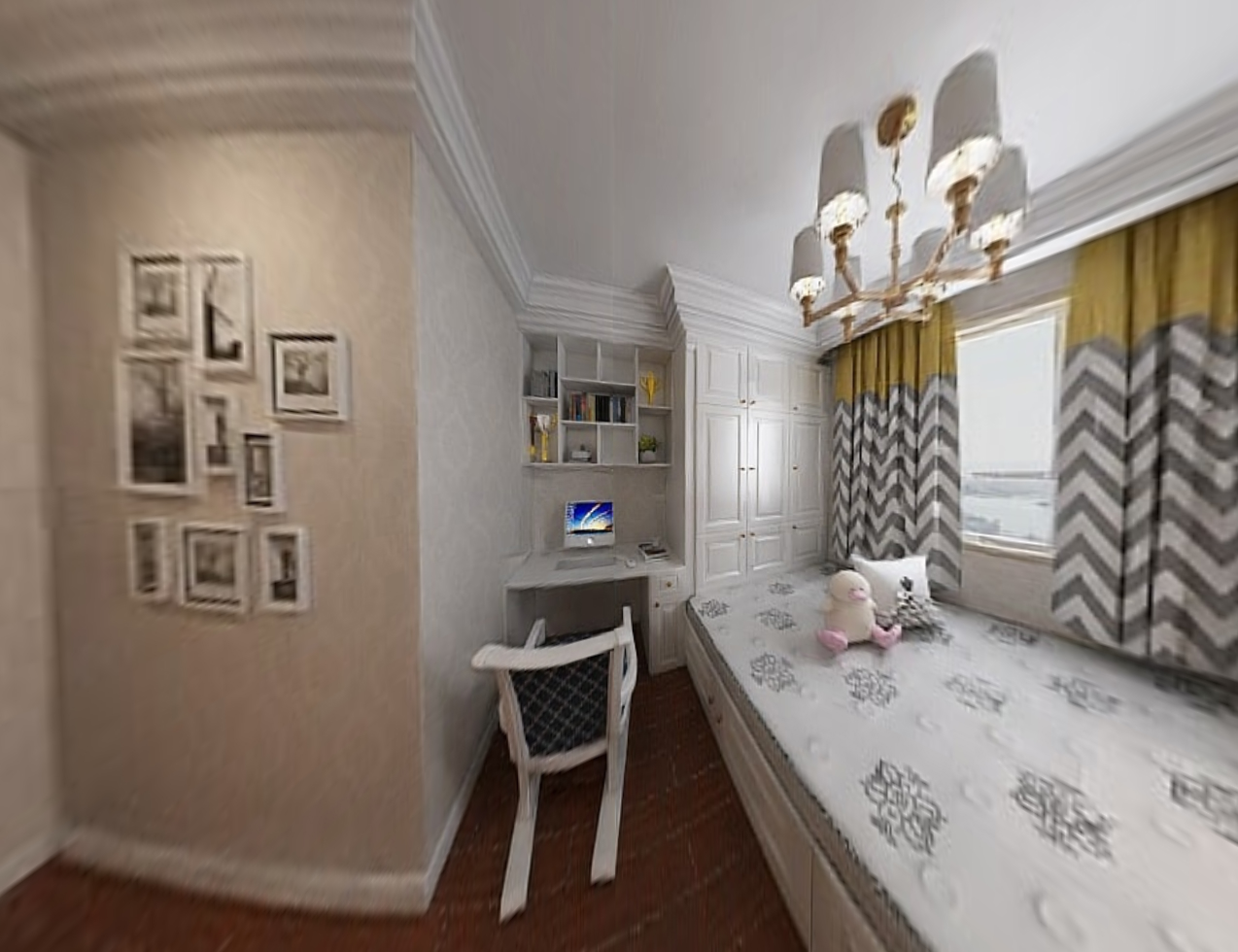}
\end{minipage}
\begin{minipage}{.208\textwidth}
\includegraphics[width=\textwidth]{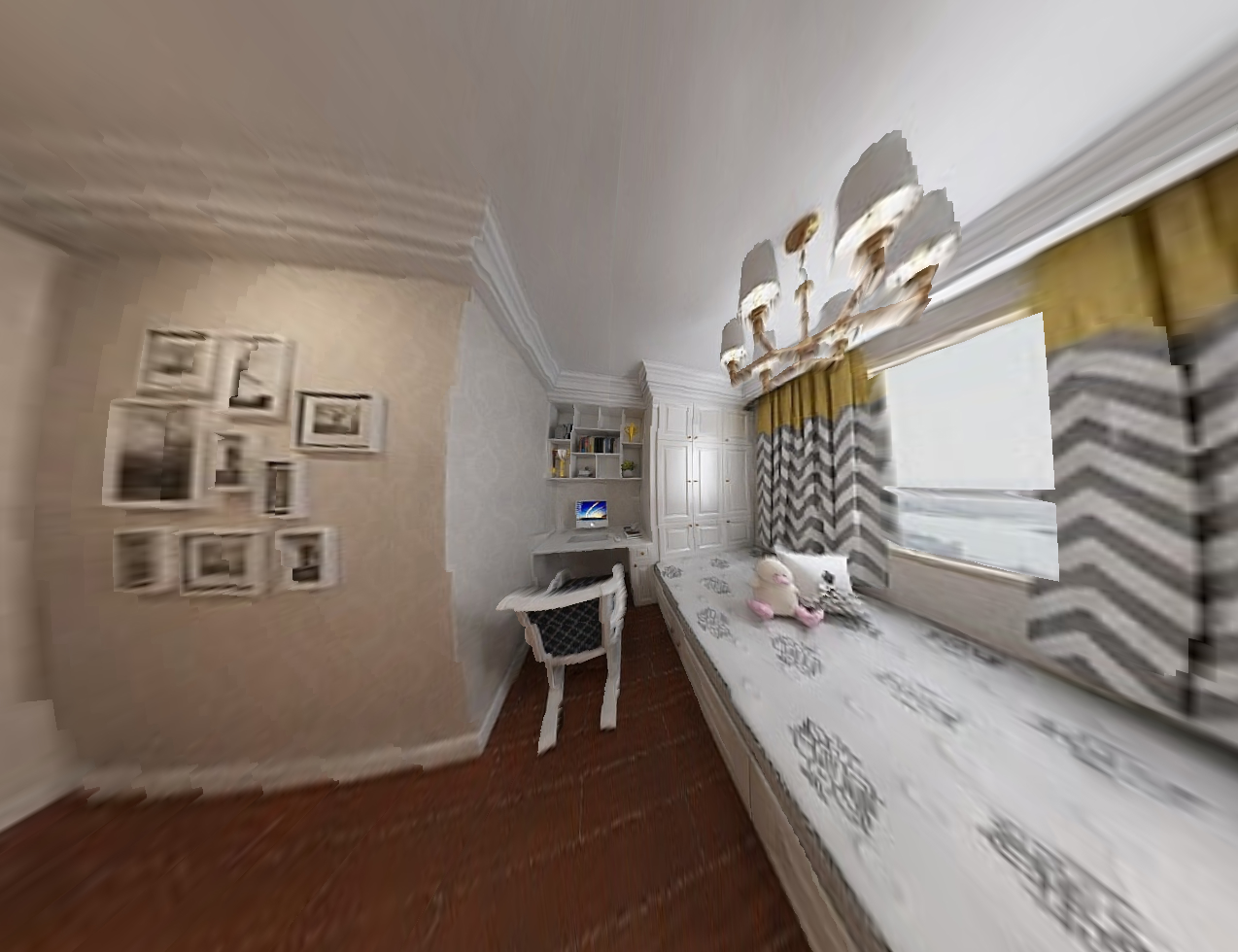}
\end{minipage}
\begin{minipage}{.208\textwidth}
\includegraphics[width=\textwidth]{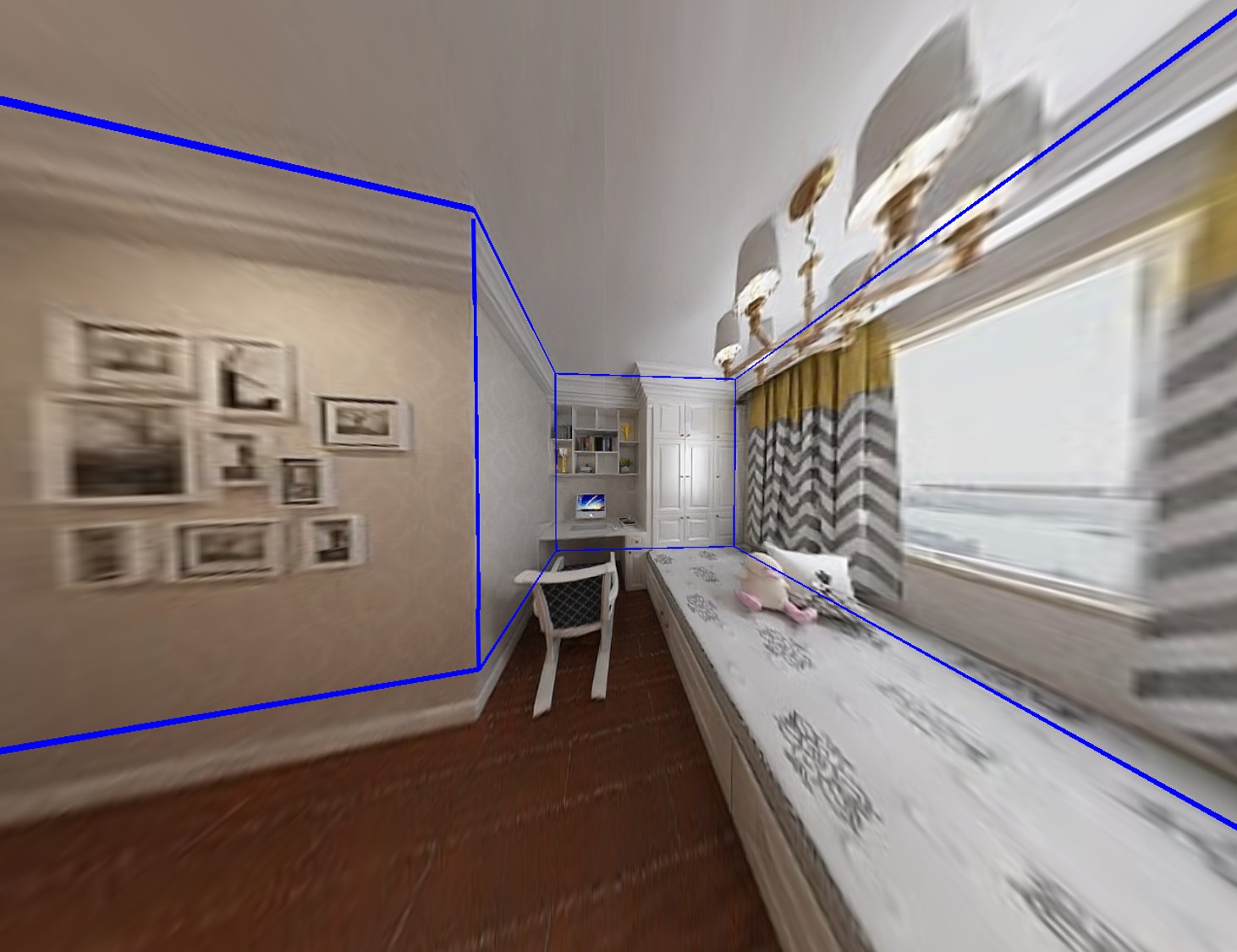}
\end{minipage}

\begin{minipage}{.15\textwidth}
\includegraphics[width=\textwidth]{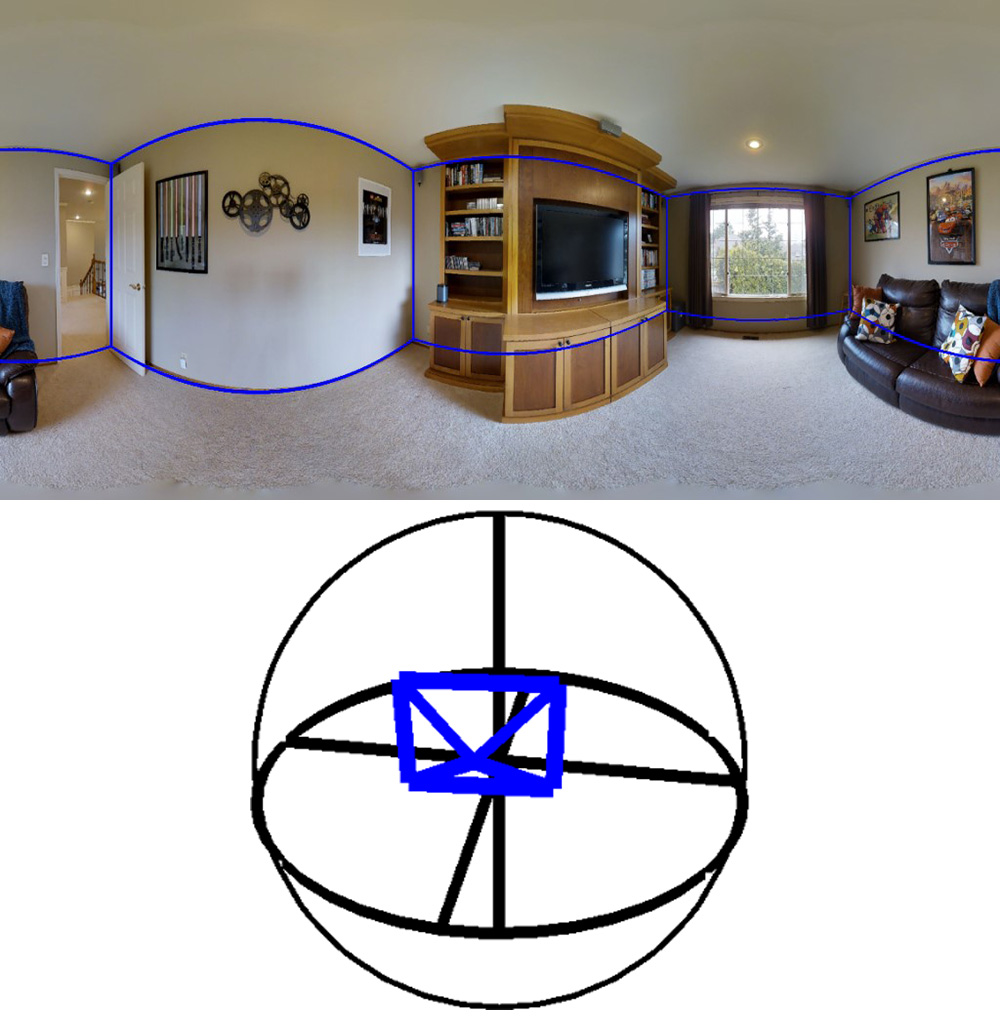}
\end{minipage}
\begin{minipage}{.208\textwidth}
\includegraphics[width=\textwidth]{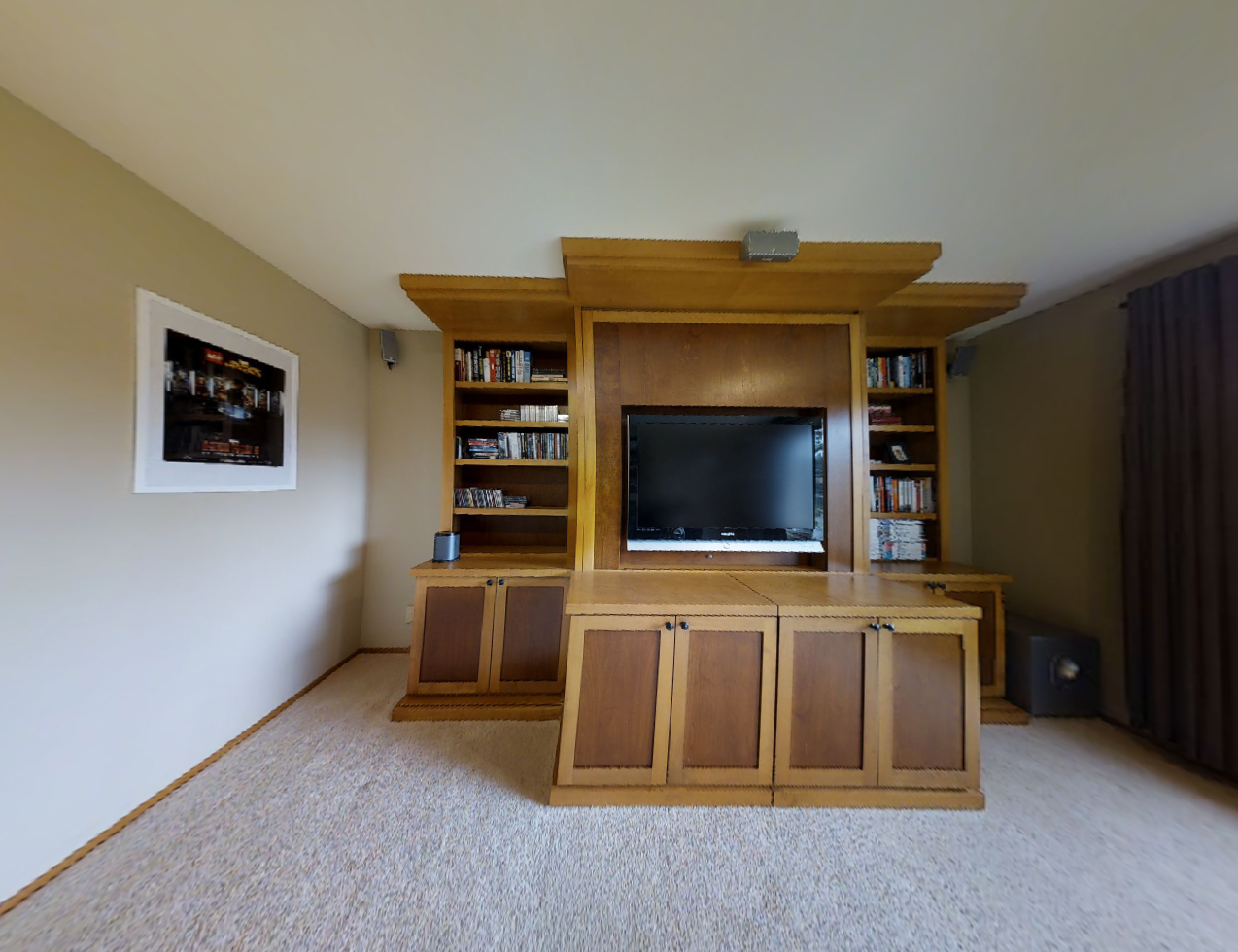}
\end{minipage}
\begin{minipage}{.208\textwidth}
\includegraphics[width=\textwidth]{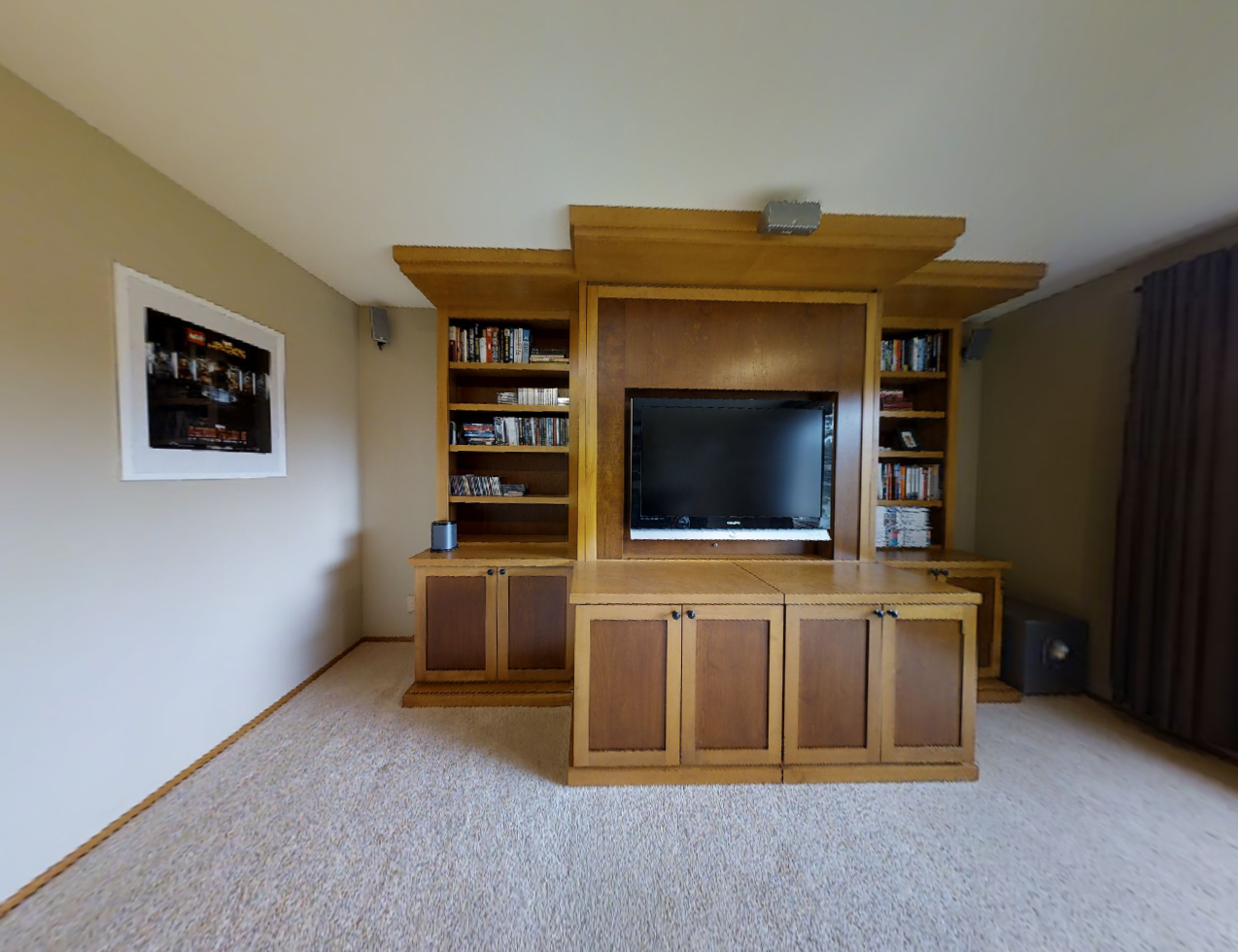}
\end{minipage}
\begin{minipage}{.208\textwidth}
\includegraphics[width=\textwidth]{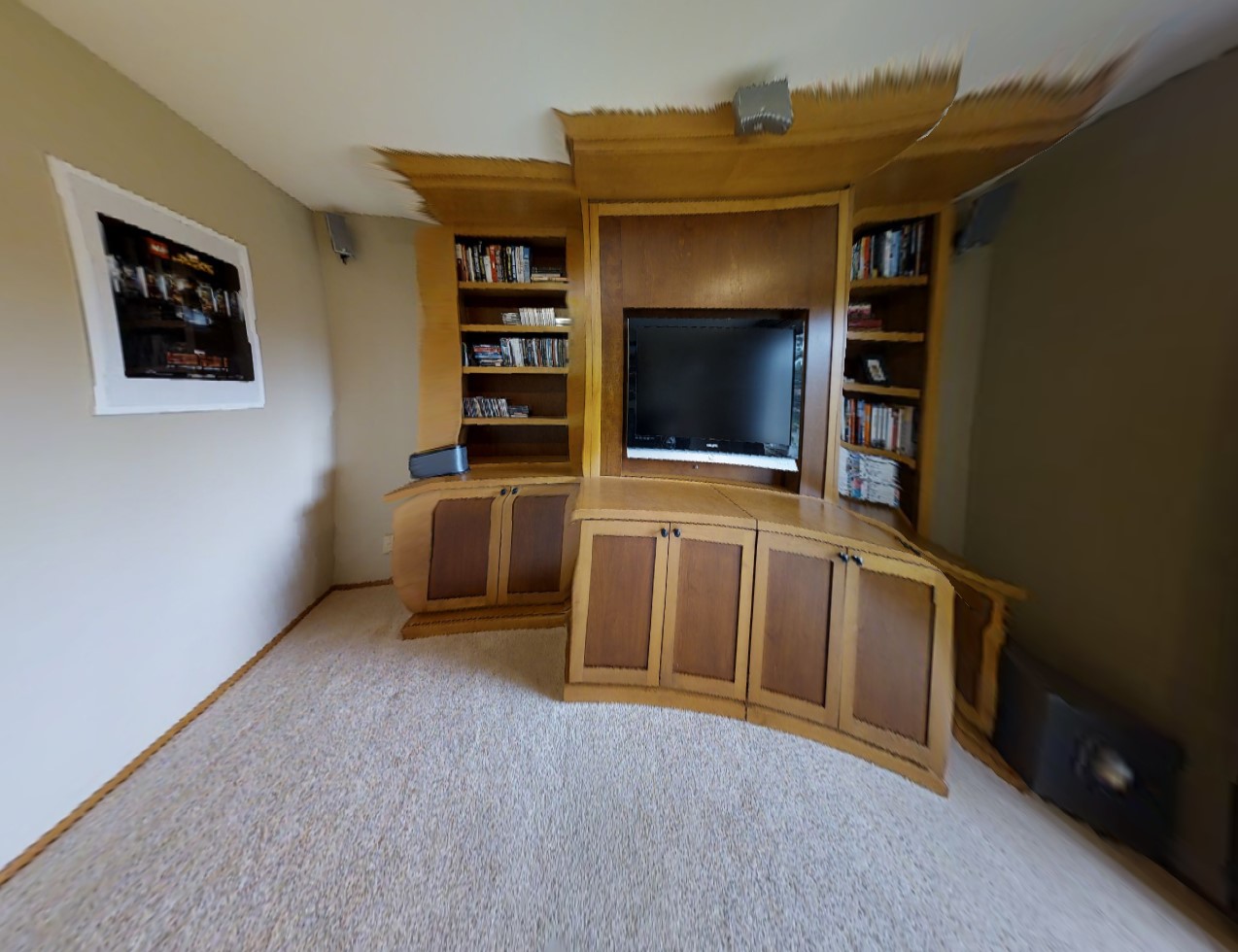}
\end{minipage}
\begin{minipage}{.208\textwidth}
\includegraphics[width=\textwidth]{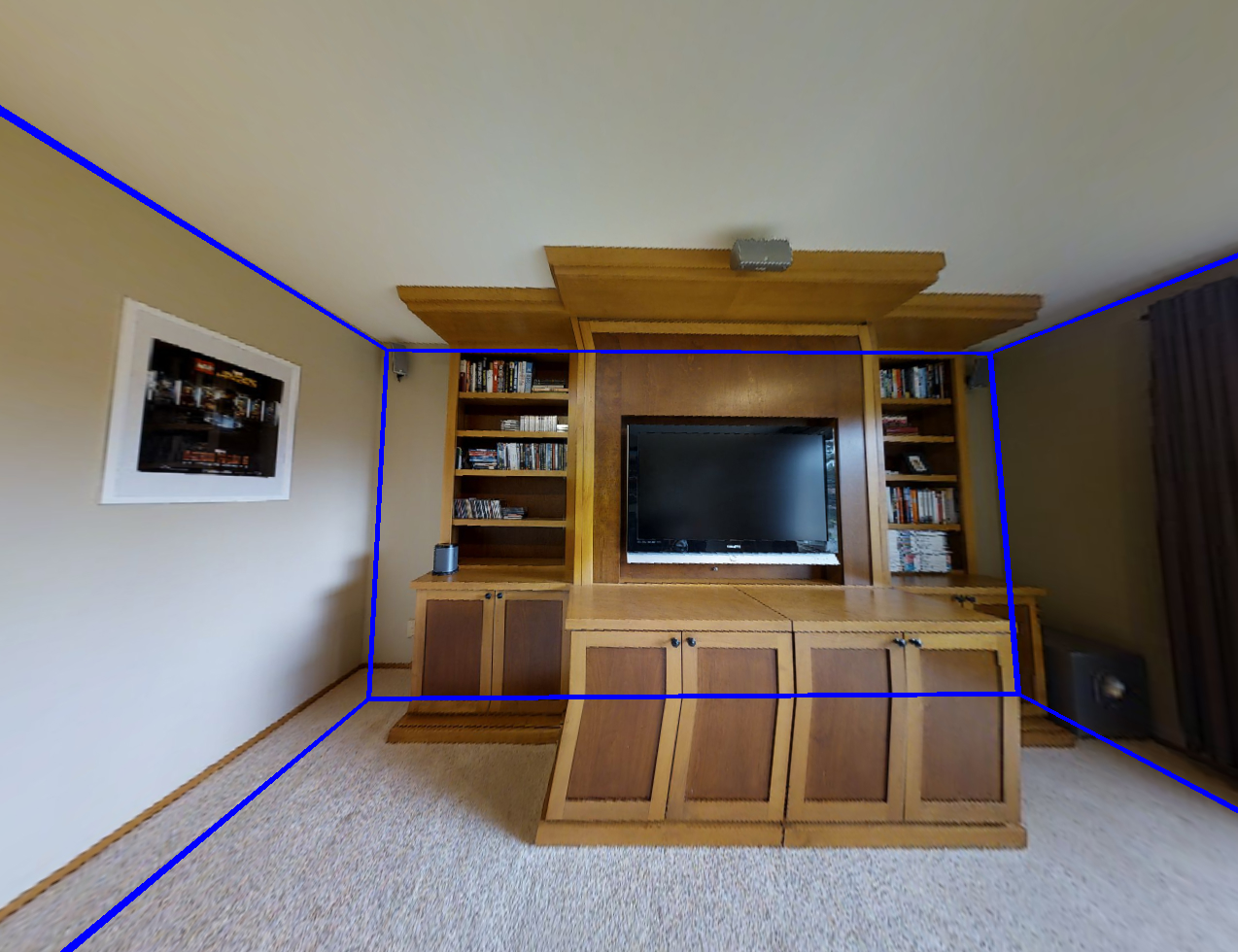}
\end{minipage}

\begin{minipage}{.15\textwidth}
\includegraphics[width=\textwidth]{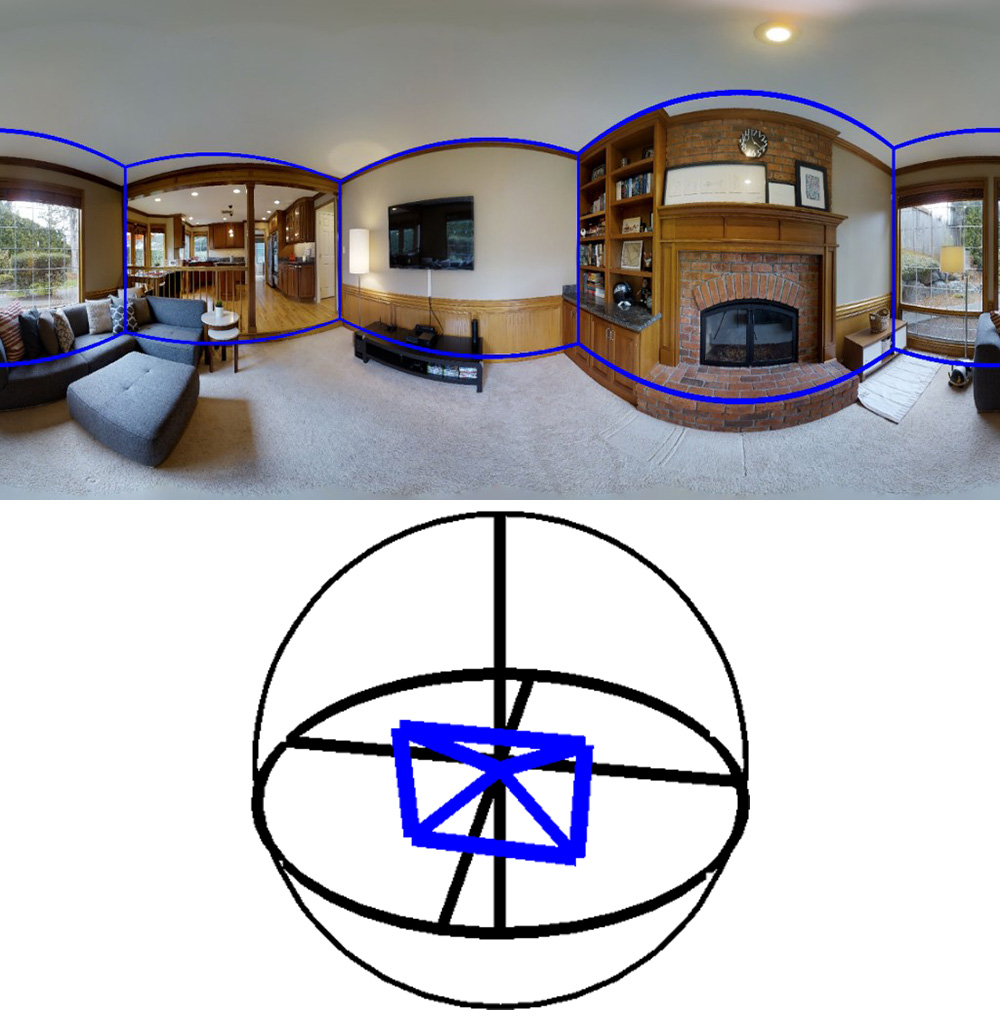}
\end{minipage}
\begin{minipage}{.208\textwidth}
\includegraphics[width=\textwidth]{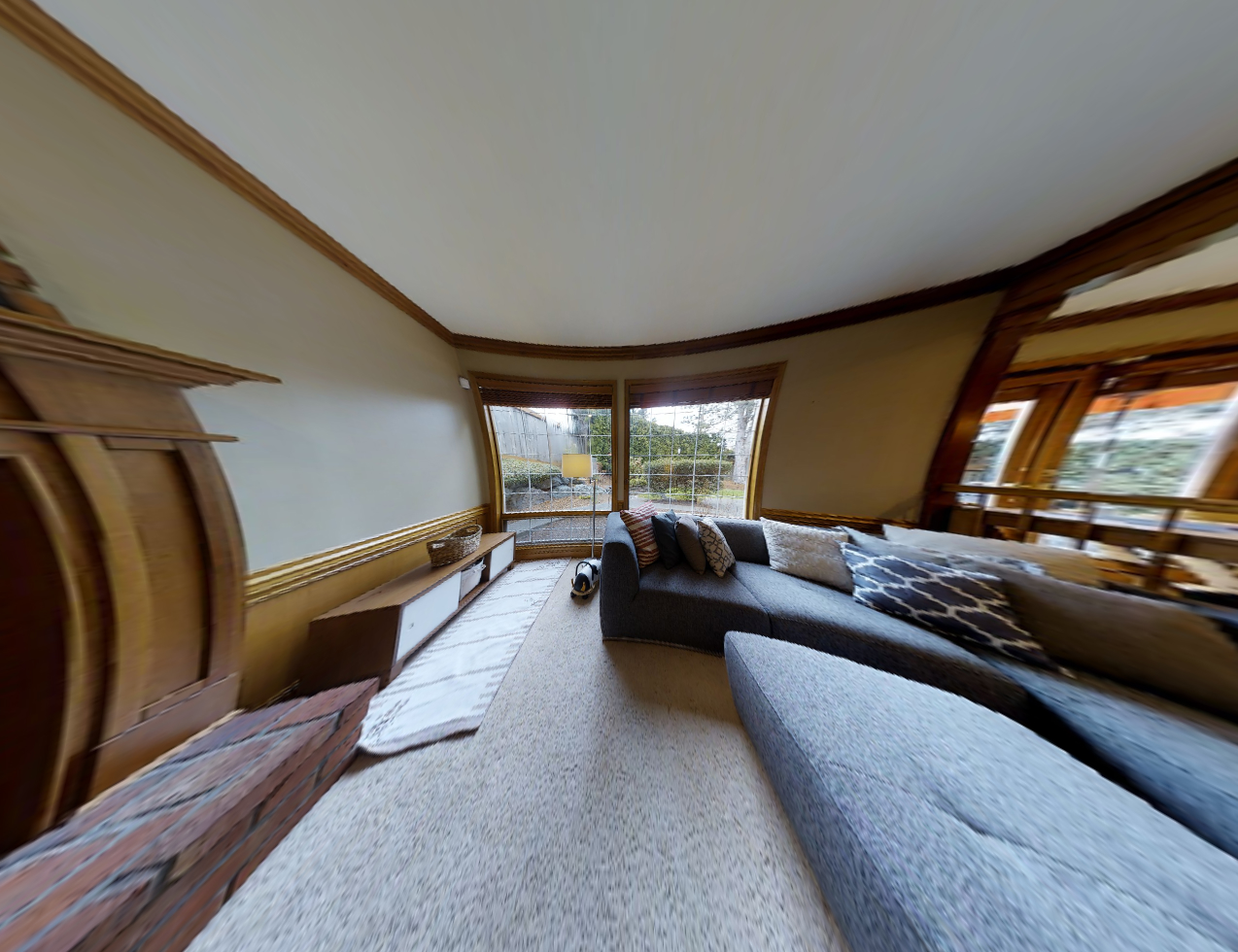}
\end{minipage}
\begin{minipage}{.208\textwidth}
\includegraphics[width=\textwidth]{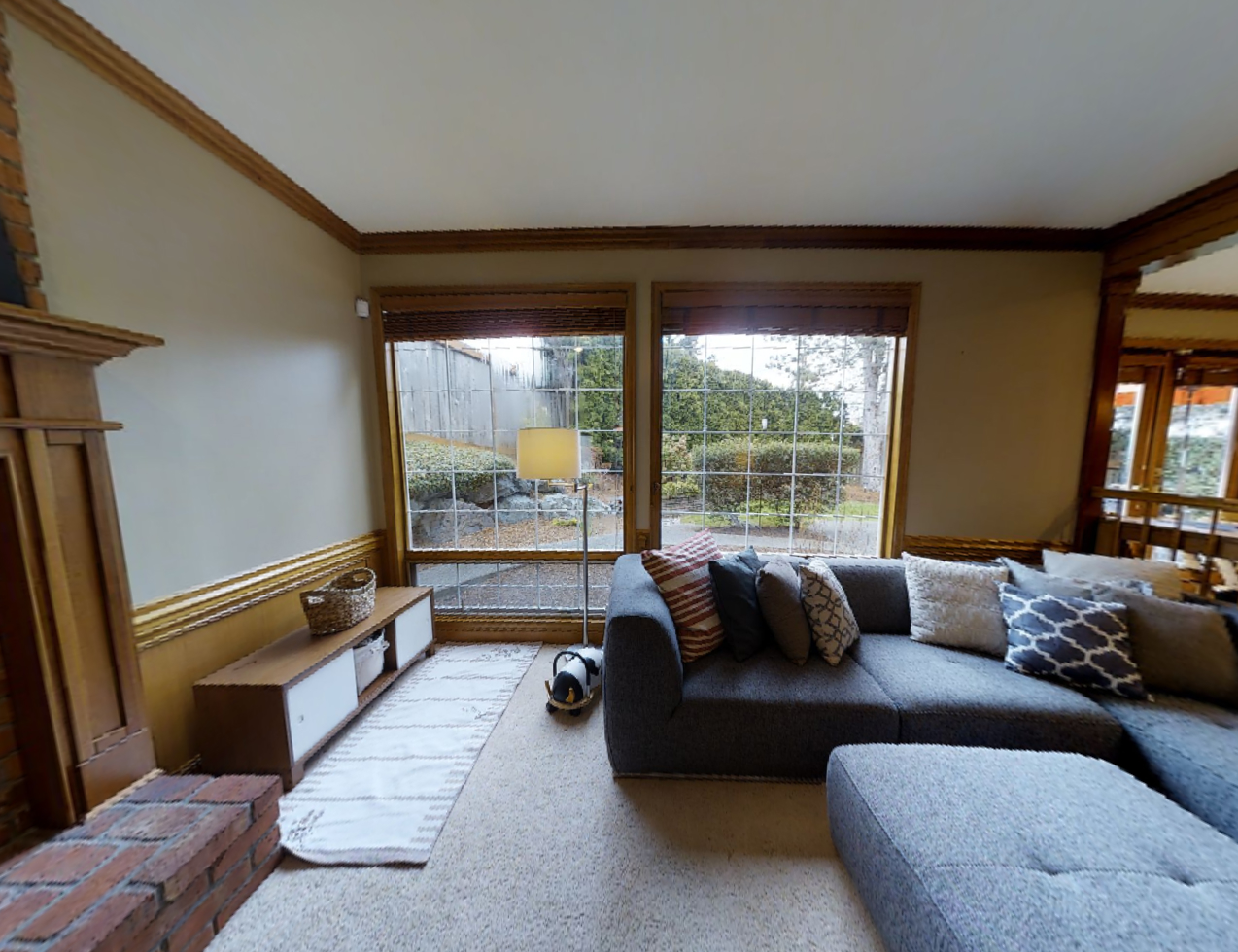}
\end{minipage}
\begin{minipage}{.208\textwidth}
\includegraphics[width=\textwidth]{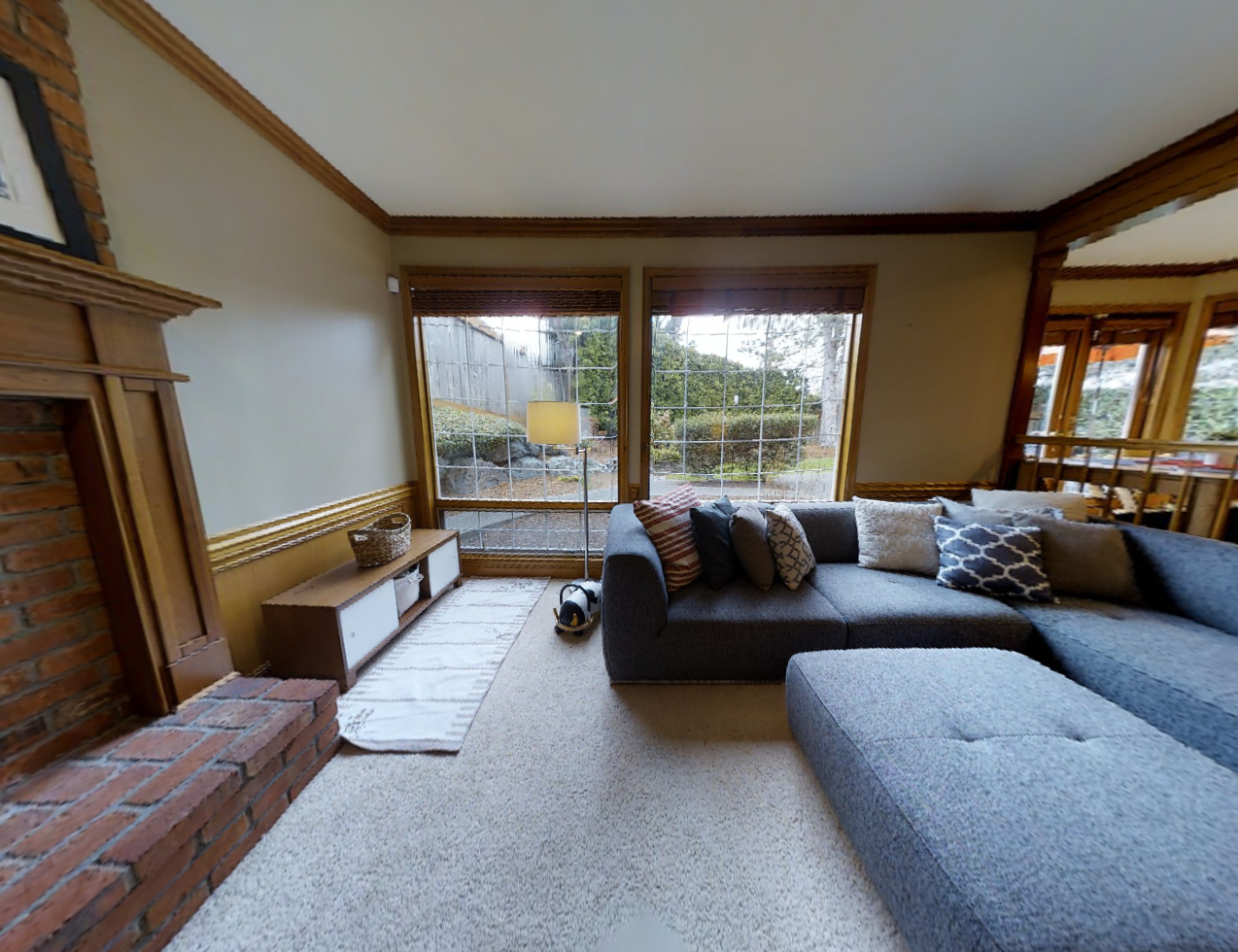}
\end{minipage}
\begin{minipage}{.208\textwidth}
\includegraphics[width=\textwidth]{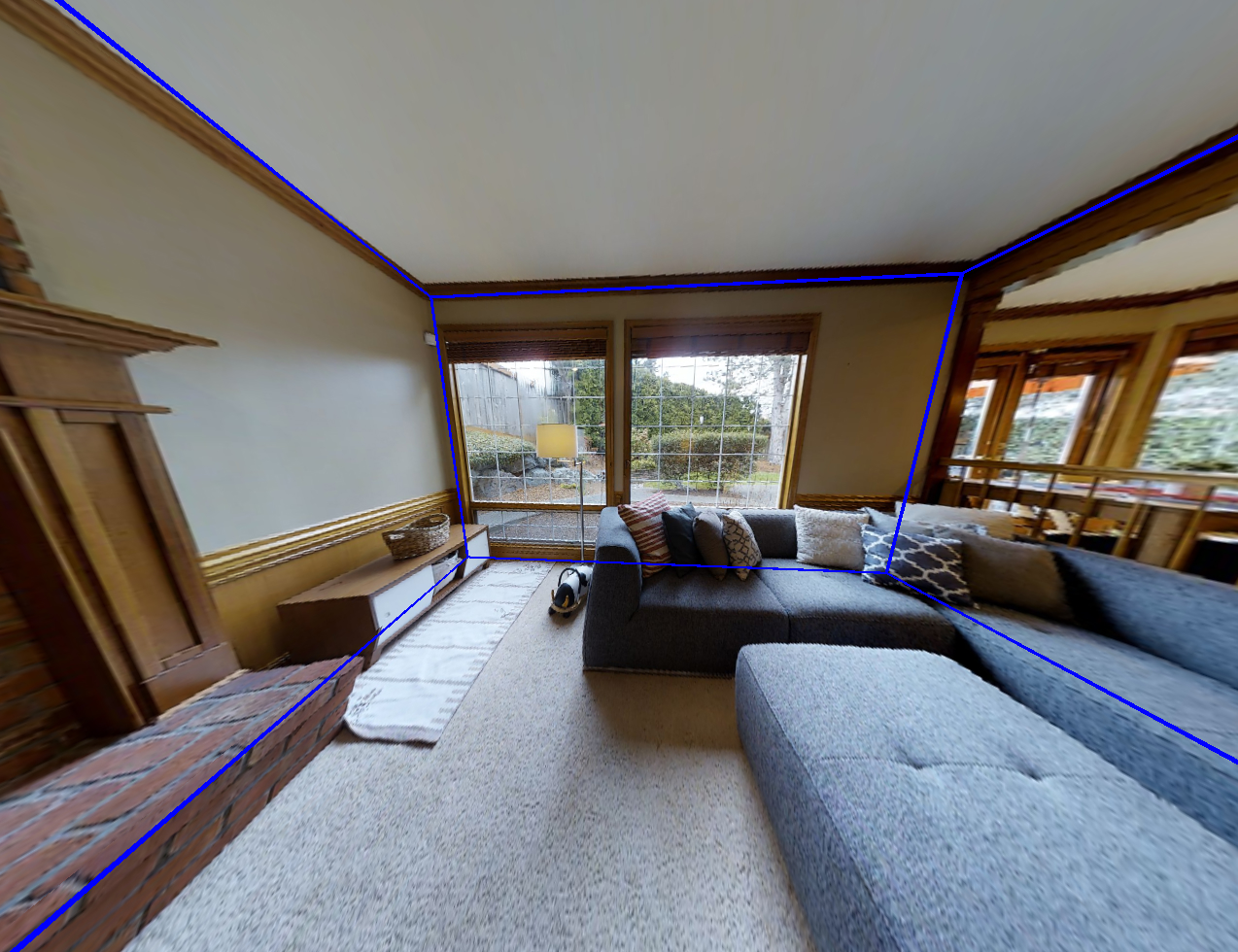}
\end{minipage}

\begin{minipage}{.15\textwidth}
\includegraphics[width=\textwidth]{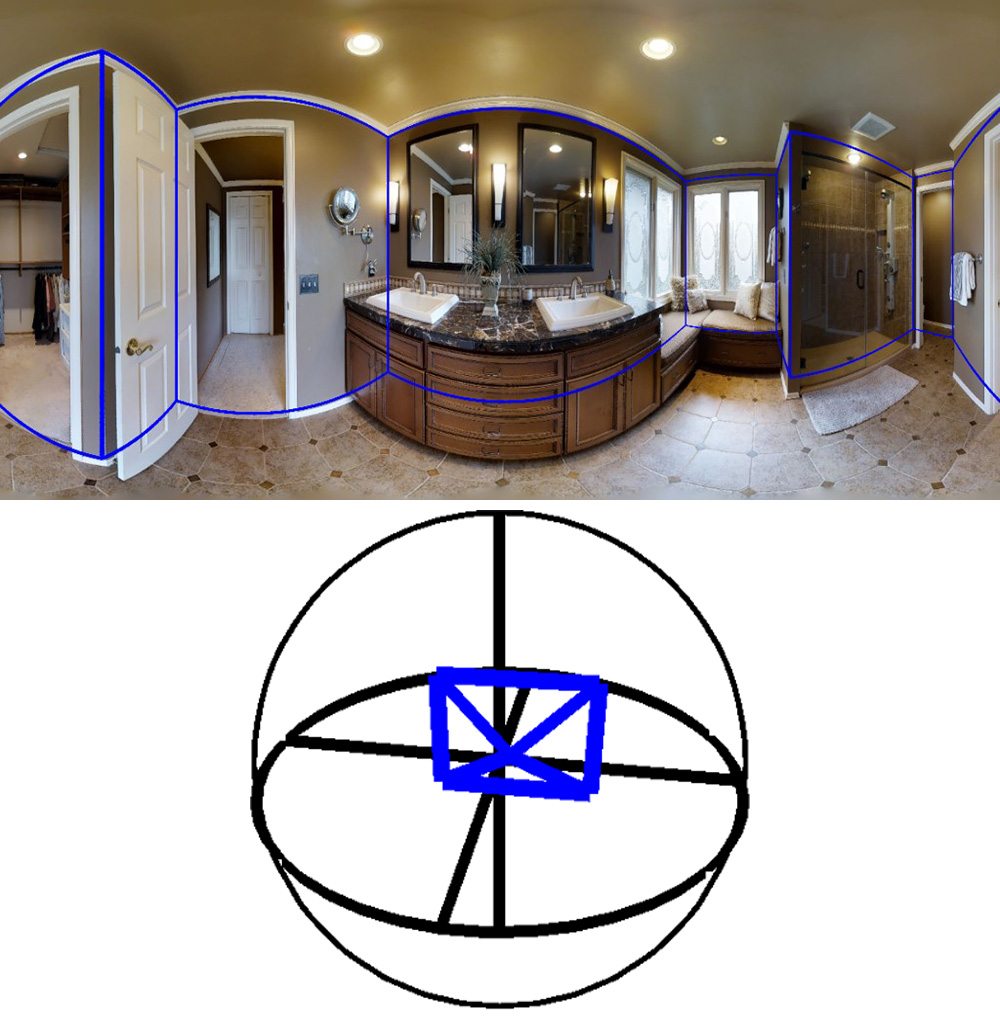}
\end{minipage}
\begin{minipage}{.208\textwidth}
\includegraphics[width=\textwidth]{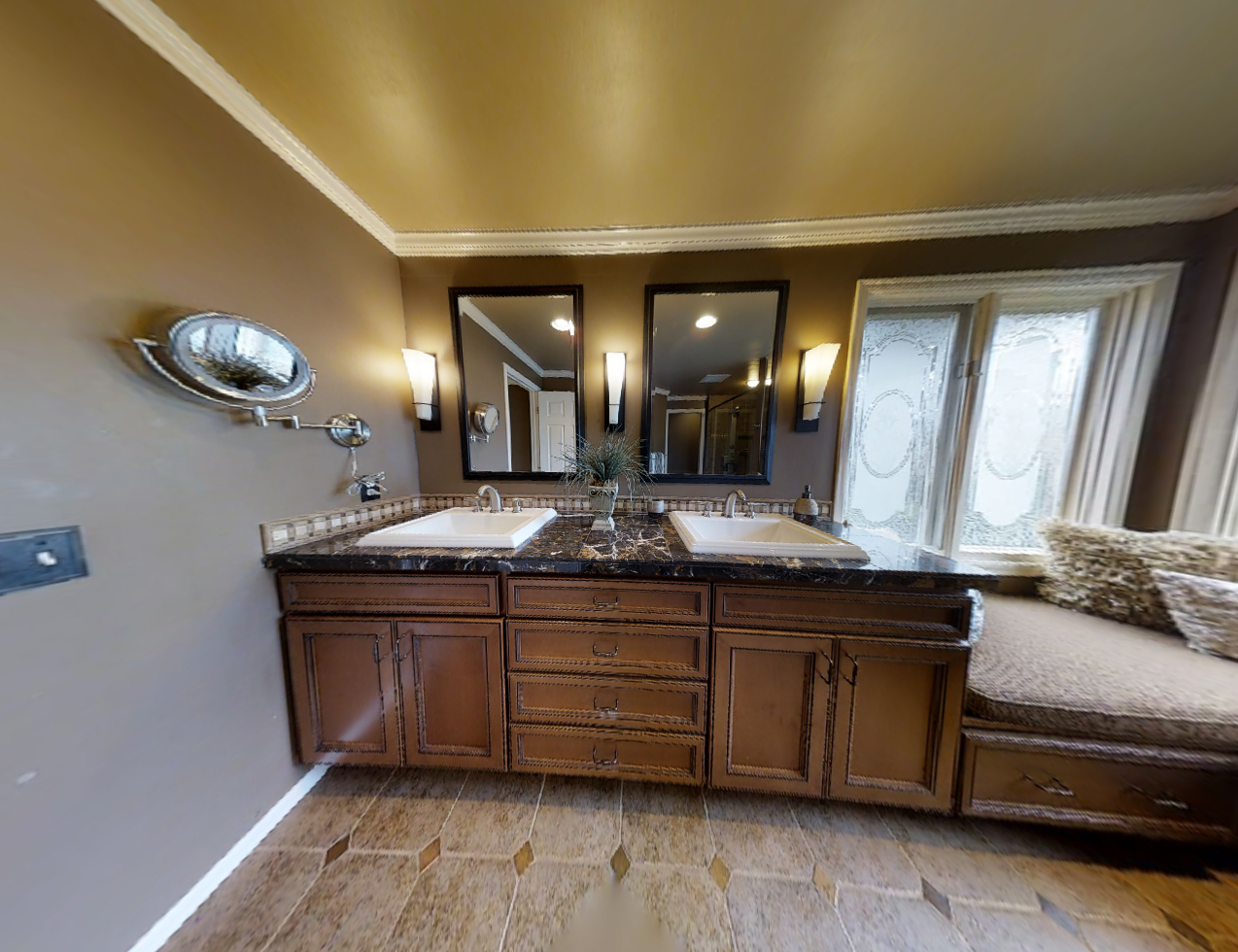}
\end{minipage}
\begin{minipage}{.208\textwidth}
\includegraphics[width=\textwidth]{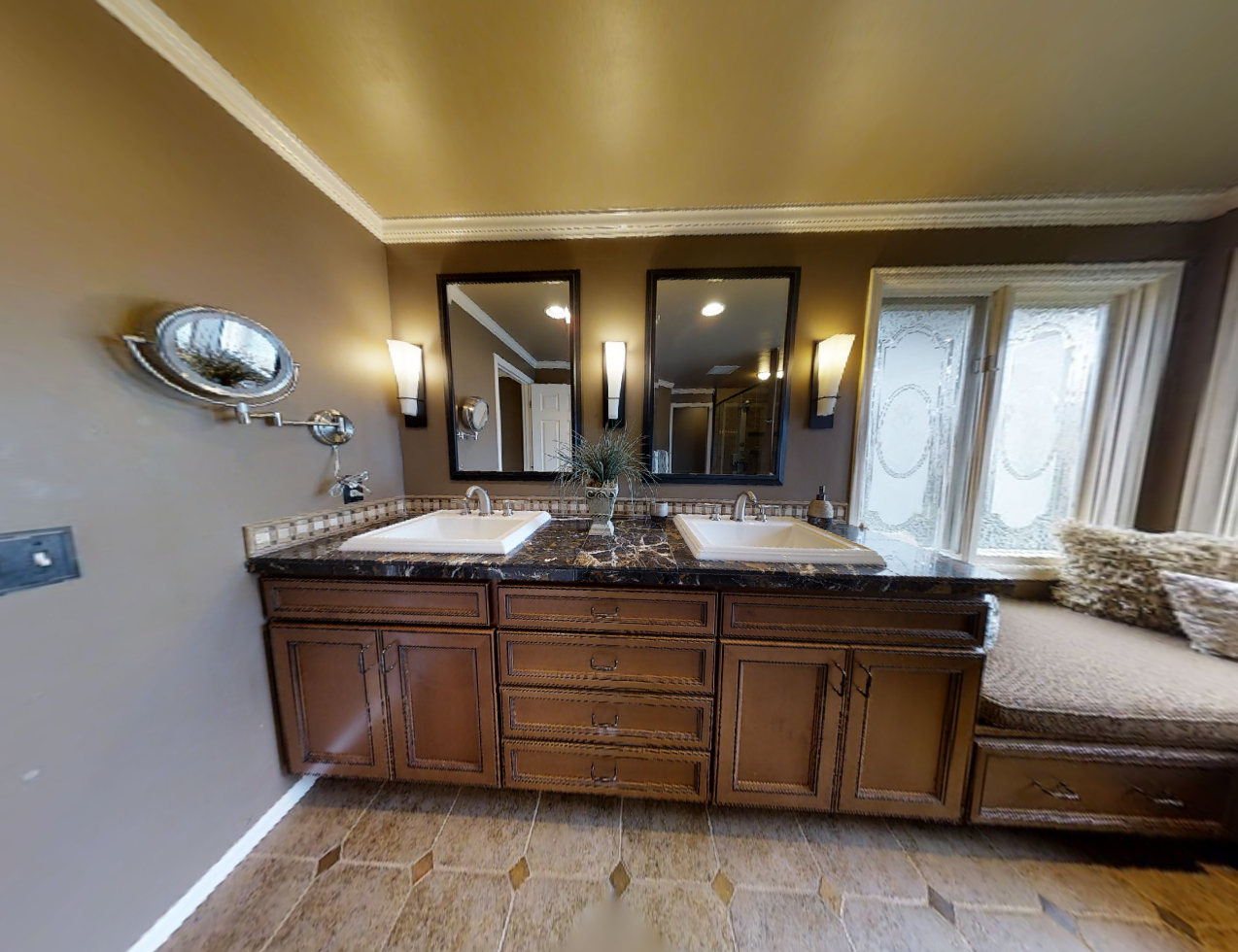}
\end{minipage}
\begin{minipage}{.208\textwidth}
\includegraphics[width=\textwidth]{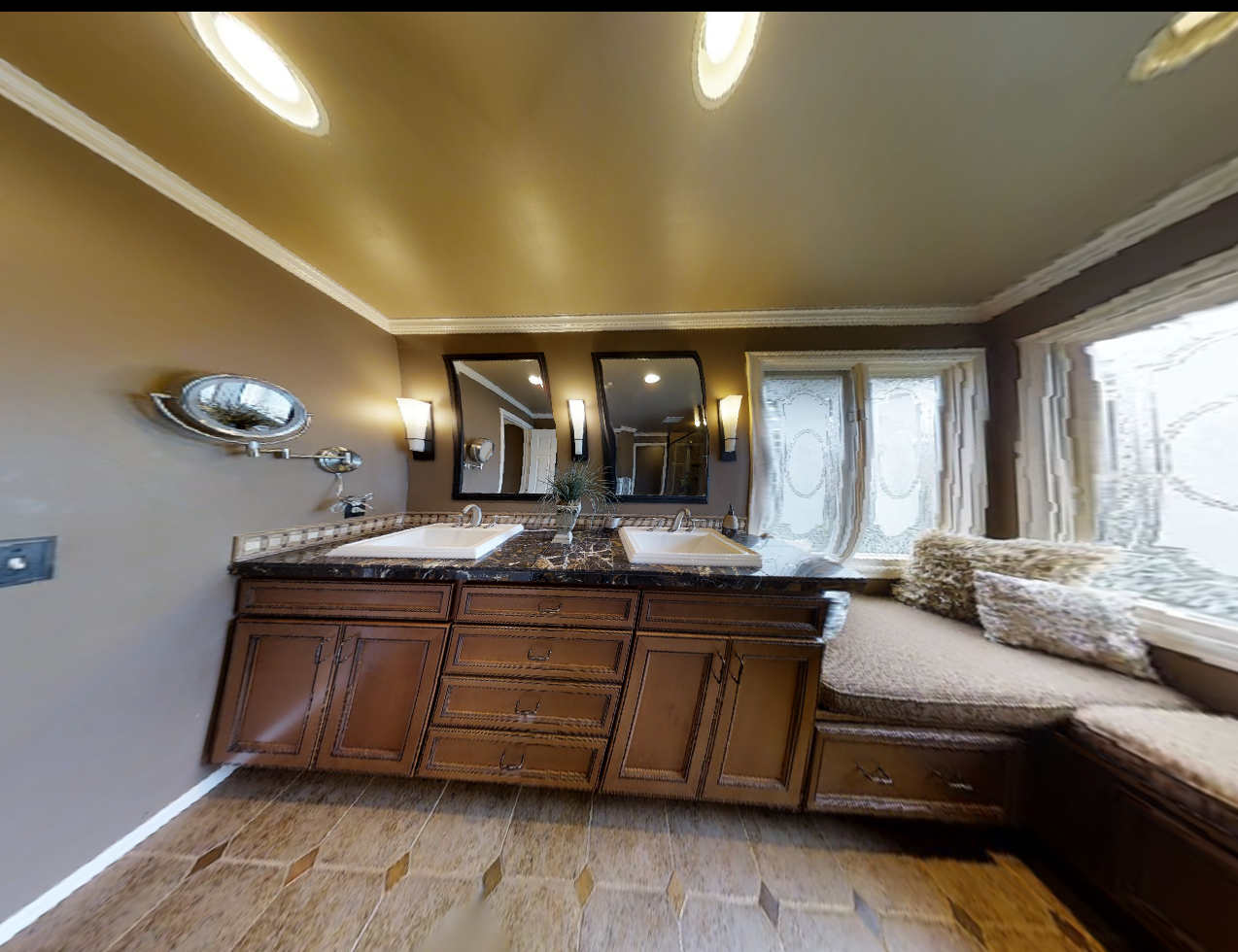}
\end{minipage}
\begin{minipage}{.208\textwidth}
\includegraphics[width=\textwidth]{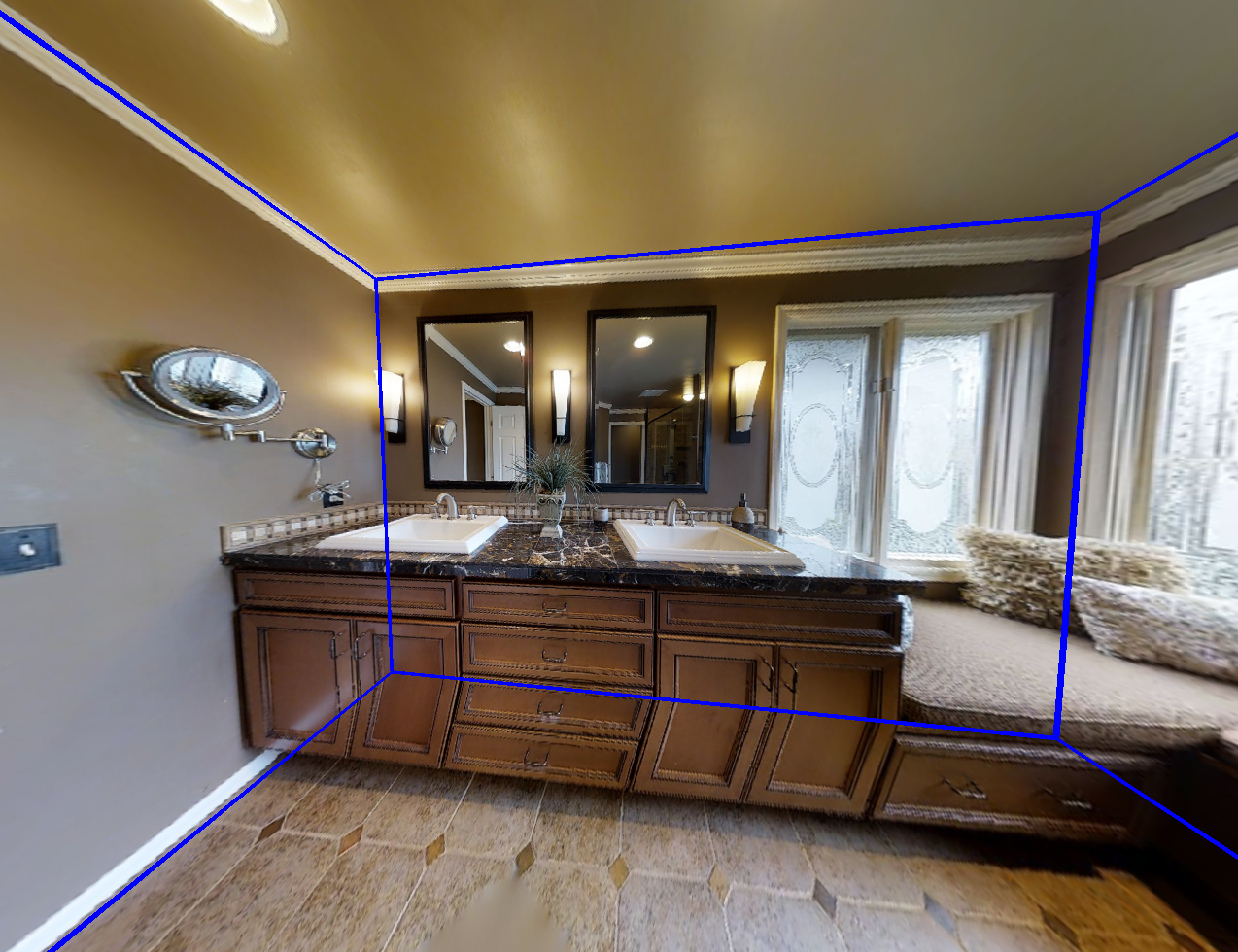}
\end{minipage}

\begin{minipage}{.15\textwidth}
\includegraphics[width=\textwidth]{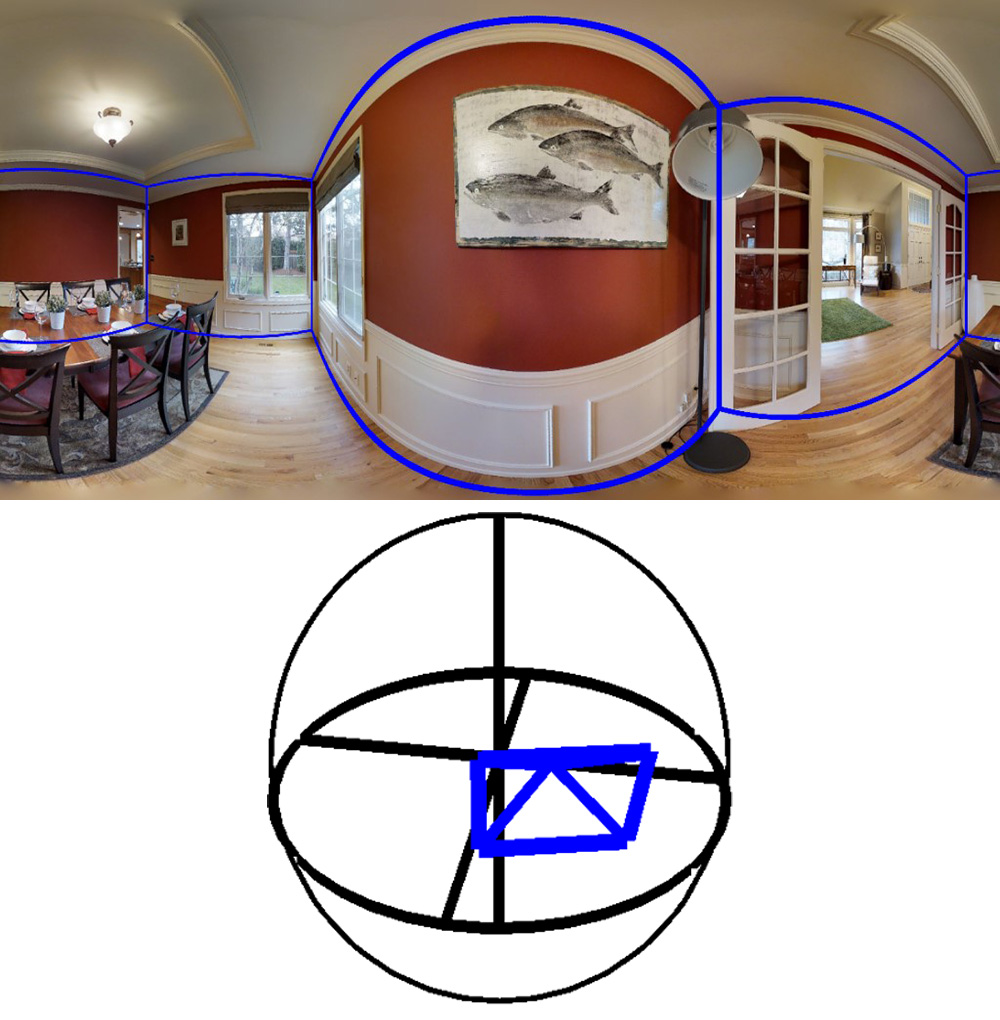}
\end{minipage}
\begin{minipage}{.208\textwidth}
\includegraphics[width=\textwidth]{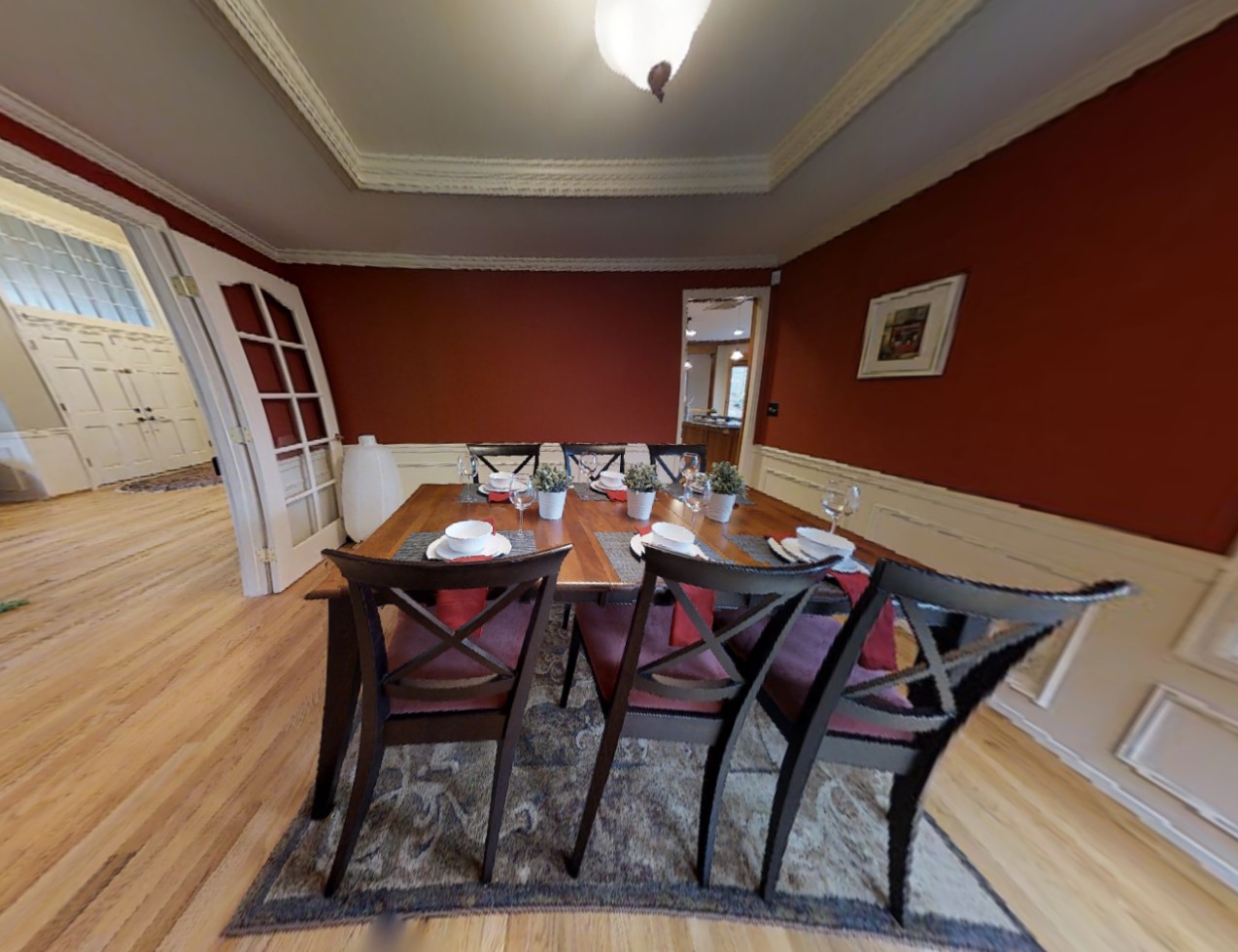}
\end{minipage}
\begin{minipage}{.208\textwidth}
\includegraphics[width=\textwidth]{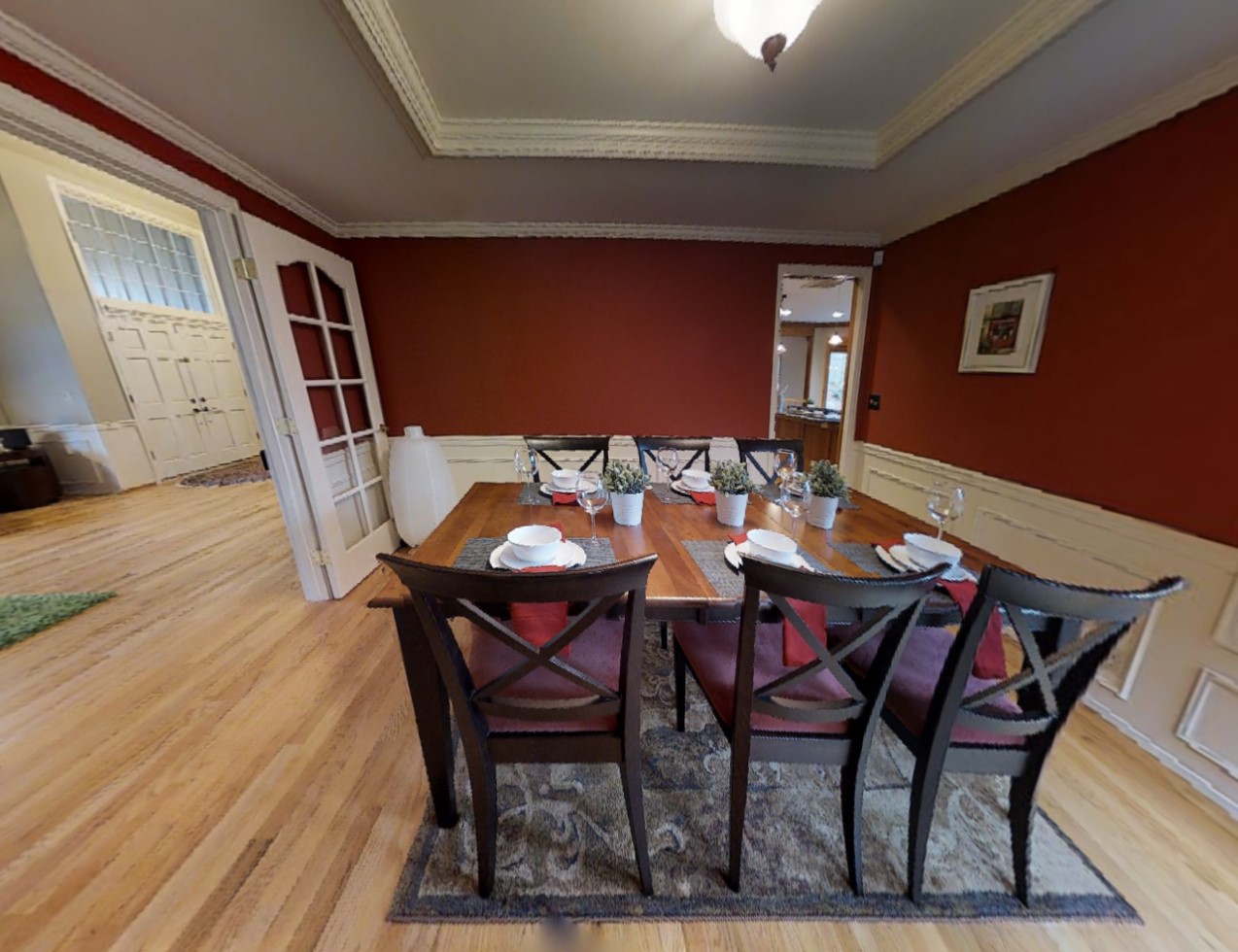}
\end{minipage}
\begin{minipage}{.208\textwidth}
\includegraphics[width=\textwidth]{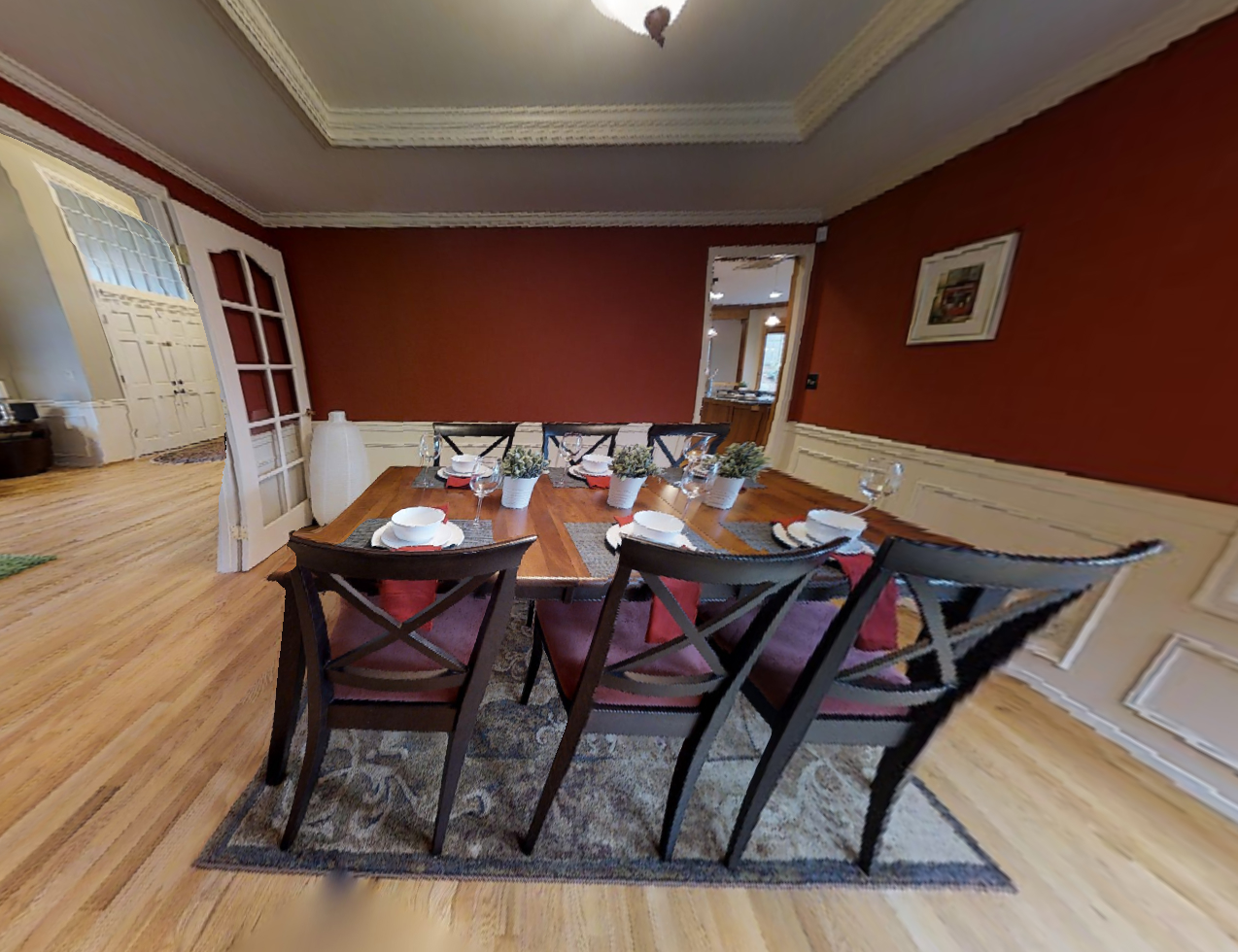}
\end{minipage}
\begin{minipage}{.208\textwidth}
\includegraphics[width=\textwidth]{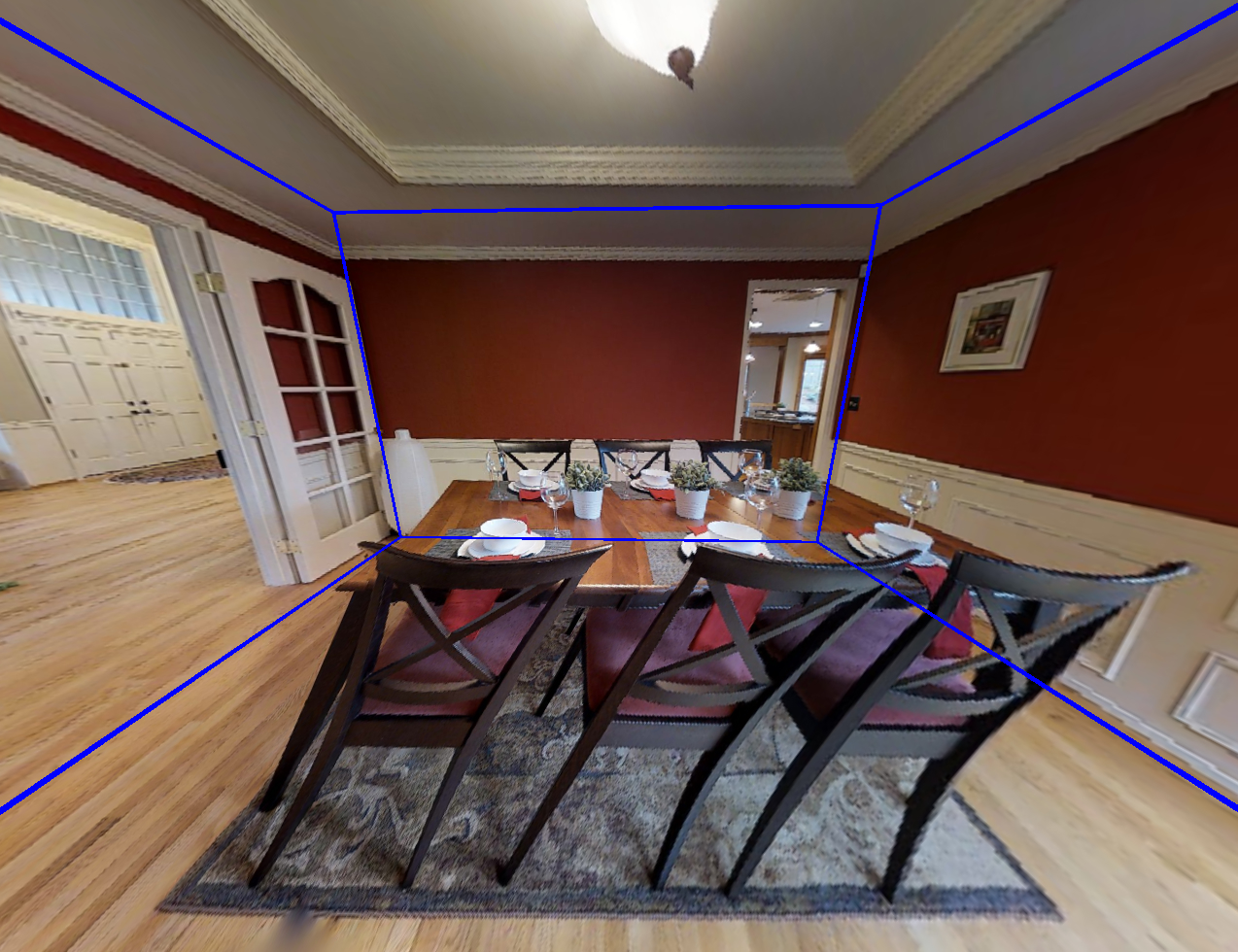}
\end{minipage}
\caption{Off-center perspective projection results. From second left to right columns: 1) by vanilla E2P projection, 2) our method, panoramas augmented with 3) depths and 4) room layouts. The first two rows are based on real-world indoor panoramas shoot by ourselves with predicted depths ("MiDaS v3.0"~\cite{Ranftl2021}) and predicted room layouts (LED2-Net~\cite{wang2021led2}). The third row is based on a panorama from the Structure3D dataset (synthetic) with ground truth depths and room layout. The last three rows are based on Gibson dataset (real-world) with predicted depths and room layouts. We show the panoramas annotated with room layouts and visualization of the off-center cameras in the left column. Bigger versions of the images can be found in the Additional Materials.}

\label{fig:generation1}
\end{figure*}

In Figure ~\ref{fig:generation1}, we compare off-center perspective projection results of our method with the vanilla E2P projection and two common approaches to augment panoramas with 3D information (per-pixel depth and room layouts). We test on panoramas from the Structure3D dataset~\cite{Structured3D} (synthetic), the Gibson dataset~\cite{xia2018gibson} (real-world), and several panoramas shoot by ourselves (we used a RICOH THETA Z1 360$^{\circ}$ camera). Note that the teaser (Figure~\ref{fig:teaser}) is also based on the Structure3D dataset. The depth information are either ground truth (available in the Structure3D dataset only) or predicted by a state-of-the-art depth prediction model ("MiDaS v3.0"~\cite{Ranftl2021}) that we found to be accurate and robust in general. We also tried another modern depth prediction model specifically for panoramas (~\cite{wang2020bifuse}) but found the results to be less accurate. The layouts are either ground truth (available in the Structure3D dataset only) or predicted by LED2-Net~\cite{wang2021led2}. In summary, our results significantly reduced the barrel distortions produced by the vanilla E2P projection. Augmenting panoramas with depths or room layouts provides more realistic 3D effects such as parallax and occlusion on-and-offs. However, glaring artifacts, such as blurs (happen when viewing "extruded" pixels by depths from sideways) and broken images features (happen when room layouts mismatch 3D objects, such as furniture, in the scene), may happen. See the accompanying video for animated versions of the results. We also provide a computer program for readers to try out on their own panoramas.

\subsection{Statistics}
\label{sec:stat}

\begin{table}
\small
\begin{center}
\begin{tabular}{| c | c c c c c|}
\hline
Method & Zeroth-q & First-q & Second-q & Third-q & Fourth-q \\
Ori. & 1.832e-05 & 0.00038 & 0.00172 & 0.00689 & 5.837e+11 \\ 
Heu. & 7.254e-15 & 0.00019 & 0.00098 & 0.00375 & 6.293e+09 \\
Opt. & 9.450e-31 & 0.00015 & 0.00071 & 0.00255 & 1.546e+06 \\
\hline
\end{tabular}
\label{tab:stat}
\caption{The zeroth- (minimum), first-, second- (median), third-, and fourth-quartiles (maximum) of the distortion values of the original camera model, the heuristic solutions, and the optimized solutions of the computational dolly-zoom effects.}
\end{center}
\end{table}

We tested on a laptop computer with 6-core 2.6GHZ CPU, 16GB ram, NVidia GTX 1650 Ti graphics card, and Windows system. We use Google Ceres-Solver to solve the computational dolly-zoom effect optimization problem. We measure the times to solve the problem in all possible camera poses (the same sampling as in Figure~\ref{fig:dolly4}). The average and largest times are 0.53 and 7 milliseconds. This means that the method is suitable for real-time applications on a reasonable computer. The computational costs of the heuristic solutions are negligible. The quartiles of the distortion values of the original cameras, heuristic solutions, and optimized solutions are shown in Table~\ref{tab:stat}. In summary, both heuristic and optimized solutions improve upon the original camera model in terms of distortion values, while the optimized solutions win over the heuristic ones by a sizeable margin.

\subsection{Limitations}
\label{sec:limitations}

Our method can be summarized as using a cylinder as the proxy mesh to define the per-pixel depths, which eliminates distortions of vertical features in 3D, and then using the computational dolly-zoom effect to find optimal alternative camera positions and FOV angles to render the same view-able regions but with minimized distortions. Our method does not create true 3D effects such as parallax and occlusion on-and-offs. Nevertheless, the illusion of a scene in 3D remains to a degree when the camera movement is small w.r.t. the depth disparity of the scene. Our hypothesis is that human brains can still deduct 3D depths by the image features (semantics, shading, straight lines, etc). Realism of the rendering begins to fray when the relative positions of the camera to the other objects significantly changed. One example is the last row in Figure~\ref{fig:xuu} (the camera moved to the other side of the table-chairs set).

\section{Conclusion and future work}
\label{sec:conclusion}

In this paper, we explore how to tackle barrel distortions in E2P projections when the camera moves away from the center. The experiments show that our two proposed approaches, namely the cylindrical projection and the computational dolly-zoom effect, significantly reduced distortions in different indoor scene datasets. The resulting renderings are smooth and intuitive, albeit lacking true 3D effects such as parallax and occlusion on-and-offs. Comparing to the state-of-the-art DL-based method~\cite{xu2021layout}, one key advantage of our method is that the resolution of perspective views is much higher. We compare our method with other methods that aim to augment panoramas with 3D information (i.e., depth information and room layouts), and show that these methods are actually more likely to produce results with noticeable artifacts.

In summary, our approach relies on novel strategies to re-sample the existing pixels in a panorama to synthesize off-center perspective views with minimized barrel distortions. For future work, we would like to expand our methods to leverage readily-available 3D information of a panorama, such as depths and room layouts predicted by neural networks. For example, creating more complex proxy meshes according to depths or layouts instead of a cylinder. More ambitious goals include creating datasets and deep-learning methods that directly synthesize novel full-resolution perspective views out of a single or a sparse set of panoramas.


\printbibliography 

\end{document}